\def\astroph{1} 

\ifnum\astroph=0
   \documentclass[manuscript]{aastex}          
\else
  \documentclass{emulateapj}
 \fi

\usepackage{epsfig,amsmath,natbib,mathrsfs}
\usepackage[dvips,usenames,dvipsnames]{color}
\usepackage{verbatim}

\newcommand{\moca}{{\sc MoCaLaTA}}
\newcommand{\lya}{Ly$\alpha$}

\definecolor{gray}{RGB}{180,180,180}

\renewcommand{\sec}[1]{Sec.~\ref{sec:#1}}
\newcommand{\fig}[1]{Fig.~\ref{fig:#1}}
\newcommand{\tab}[1]{Tab.~\ref{tab:#1}}

\newcommand{\Fig}[1]{Figure \ref{fig:#1}}

\renewcommand\ion[2]{#1{\sc #2}}
\def\HI{\mbox{\ion{H}{i}}}
\def\HII{\mbox{\ion{H}{ii}}}
\newcommand{\Pcl}{\ensuremath{P_{\mathrm{cl}}}}
\newcommand{\nHI}   {\ensuremath{n_{\textrm{{\scriptsize H}{\tiny \hspace{.1mm}I}}}}}
\newcommand{\NHI}   {\ensuremath{N_{\textrm{{\scriptsize H}{\tiny \hspace{.1mm}I}}}}}
\newcommand{\NHIcl} {\ensuremath{N_{\textrm{{\scriptsize H}{\tiny \hspace{.1mm}I},cl}}}}
\newcommand{\nHIcl} {\ensuremath{n_{\textrm{{\scriptsize H}{\tiny \hspace{.1mm}I},cl}}}}
\newcommand{\NHIICM}{\ensuremath{N_{\textrm{{\scriptsize H}{\tiny \hspace{.1mm}I},ICM}}}}
\newcommand{\nHIICM}{\ensuremath{n_{\textrm{{\scriptsize H}{\tiny \hspace{.1mm}I},ICM}}}}
\newcommand{\nHICM} {\ensuremath{n_{\textrm{H,ICM}}}}
\newcommand{\xHIICM}{\ensuremath{x_{\textrm{{\scriptsize H}{\tiny \hspace{.1mm}I},ICM}}}}
\newcommand{\xHI}   {\ensuremath{x_{\textrm{{\scriptsize H}{\tiny \hspace{.1mm}I}}}}}

\newcommand{\xid}   {\ensuremath{\xi_{\textrm{d}}}}
\newcommand{\xidcl} {\ensuremath{\xi_{\textrm{d,cl}}}}
\newcommand{\xidICM}{\ensuremath{\xi_{\textrm{d,ICM}}}}
\newcommand{\kms}{\ensuremath{{\rm km\,s^{-1}}}}

\def\ergscm2{\mbox{\,erg~s$^{-1}$~cm$^{-2}$}}
\def\pcms{\mbox{cm$^{-2}$}} 
\def\pcmc{\mbox{cm$^{-3}$}} 
\def\cm2{\mbox{cm$^{2}$}}
\newcommand{\nd}{\ensuremath{n_{\textrm{d}}}}
\newcommand{\ave}[1]{\ensuremath{\langle #1 \rangle}}
\newcommand{\ten}[1]{\ensuremath{10^{#1}}}
\newcommand{\e}[1]  {\ensuremath{\times10^{#1}}}
\def\fc{\mbox{$f_{\mathrm{c}}$}}
\def\sigV{\mbox{$\sigma_{V,\textrm{cl}}$}}
\def\Vout{\mbox{$V_{\mathrm{out}}$}}
\def\ebv{\mbox{$E(B-V)$}}
\def\Msun{\mbox{$M_\odot$}}
\def\Zsun{\mbox{$Z_\odot$}}
\defcitealias{neu91}{N91}
\defcitealias{han06}{HO06}
\newcommand{\neu}{\citetalias{neu91}}
\newcommand{\han}{\citetalias{han06}}
\renewcommand{\P}{\ensuremath{\mathscr{P}}} 

\newcommand{\ewint}{\ensuremath{W_{\mathrm{int}}}}
\newcommand{\ewem} {\ensuremath{W_{\mathrm{em}}}}
\newcommand{\ewobs}{\ensuremath{W_{\mathrm{obs}}}}
\newcommand{\rgal}{\ensuremath{r_{\mathrm{gal}}}}
\newcommand{\rcl}{\ensuremath{r_{\mathrm{cl}}}}
\newcommand{\taud}{\ensuremath{\langle\tau_{\mathrm{d}}\rangle}}
\newcommand{\taua}{\ensuremath{\langle\tau_{\mathrm{a}}\rangle}}
\newcommand{\dtot}{\ensuremath{\langle d_{\mathrm{tot}}\rangle}}

\definecolor{DarkRed}{RGB}{195,0,0} 

\begin{document}
%
\shorttitle{On the (non)-enhancement of the \lya\ EW by a multiphase ISM}
\shortauthors{Laursen et al.}
\title{On the (non-)enhancement of the \lya\ equivalent width\\
  by a multiphase interstellar medium}
\author{Peter Laursen\altaffilmark{1,2},
        Florent Duval\altaffilmark{1}, \&
        G\"oran \"Ostlin\altaffilmark{1}}
\altaffiltext{1}{The Oskar Klein Centre, Dept.~of Astronomy, AlbaNova,
  Stockholm University, SE-10691 Stockholm, Sweden; email: plaur@astro.su.se.}
\altaffiltext{2}{Dark Cosmology Centre, Niels Bohr Institute, University of
 Copenhagen, Juliane Maries Vej~30, DK-2100, Copenhagen {\O}, Denmark.}

\begin{abstract}
It has been suggested that radiative transfer effects may explain the unusually
high equivalent widths (EWs) of the \lya\ line, observed occasionally from
starburst galaxies, especially at high redshifts.
If the dust is locked up inside high-density clouds dispersed in an empty
intercloud medium, the \lya\ photons could scatter off of the surfaces of
the clouds, effectively having their journey confined to the dustless medium.
The continuum radiation, on the other hand, does not scatter, and would thus be
subject to absorption inside the clouds.

This scenario is routinely invoked when \lya\ EWs higher than what is expected
theoretically are observed, although the ideal conditions under which the
results are derived usually are not considered.
Here we systematically examine the relevant physical parameters in this
idealized framework, testing whether any astrophysically realistic scenarios
may lead to such an effect. It is found that although clumpiness indeed
facilitates the escape of \lya, it is highly unlikely that any real
interstellar media should result in a preferential escape of \lya\ over
continuum radiation. Other possible causes are discussed, and it is concluded
that the observed high EWs are more likely to be caused by cooling
radiation from cold accretion and/or anisotropic escape of the \lya\ radiation.
\end{abstract}



\section{Introduction}

Our understanding of the high-redshift Universe has expanded tremendously
during the last decade, especially due a single specific emission line, the
\lya\ line.  Tracing in particular galaxies in the process of forming, the
shape, intensity, and spatial distribution of this line carry a wealth of
information. However, while our theoretical understanding of the physical
processes that govern the radiative transfer (RT) of \lya\ has also advanced
significantly, it is evident that the complexity of density fields, gas
kinematics, dust distribution, line formation, impact of the intergalactic
medium (IGM), etc.\ render individually observed lines quite difficult to
interpret in detail.

As \lya\ emitting galaxies (LAEs) generally are relatively small
\citep{gaw06,lai07,nil07} of low (specific) star formation rate
\citep{fyn03,nil11}, our lack of knowledge is in part due to the difficulties
in observing faint galaxies. This obstacle may to some extend be overcome
using gravitationally lensed LAEs \citep[e.g.][]{fos03,qui09,chr12a,chr12b},
which may boost the observed flux by more than an order of magnitude. Larger
surveys of the statistical properties of LAEs, such as their luminosity
function \citep[LF,][]{ouc05,ouc08,shi06,kas06,daw07,gro09} and clustering properties
\citep{gaw07,gro07} may overcome some of these complications, although still
our ignorance of RT processes easily may result in systematic errors. A common
technique by which to assess the stellar population of individual galaxies at
high redshifts is to measure the equivalent width (EW) $W$ of the \lya\ line
\citep[e.g.][]{gro07,ouc08,sta10}.
This quantity is defined as the ratio of the integrated line flux to continuum
flux density, $W = \int\! d\lambda\,[F(\lambda)\!-\!F_0]/\!F_0$, where the
exact limits of the integral are not important, as long as the full line is
included. For sloped continua, the wavelength dependency of the continuum must
also be taken into account. For high EWs, $W$ can be approximated by the
relative escape fraction of \lya\ and continuum photons. The EW depends on
galactic parameters such as the initial mass function (IMF) and the gas
metallicity \citep[e.g.][]{sch02}, and can consequently be used as a probe of
these quantities.

In young galaxies, \lya\ is produced mainly from recombinations following the
ionization of hydrogen surrounding O and B stars. Since these stars are
short-lived, a few Myr after an initial starburst the EW will generally
decrease significantly. Stellar population syntheses
\citep[e.g.][]{cha93,val93,sch03}
predict that the EW should initially reach $\sim 240$ {\AA}, eventually
declining and settling on roughly 80 {\AA}, where the exact values depend on
the assumed IMF and metallicity of the population. Curiously, several
observational studies have reported on the detection of much higher EWs
\citep[e.g.][]{hu96,kud00,mal02,rho03,daw04,hu04,shi06,gro07,ouc08,nil09,kas11,
kas12}.
While some of these extreme EW galaxies can be attributed to AGN activity,
in many cases this explanation is explicitly excluded. Moreover, such high EWs
are not readily explained by simply assuming a more top-heavy IMF. Indeed, the
observations pose a serious challenge to our understanding of both galaxy
formation, stellar evolution, and radiative transfer.

In this work, we will distinguish between three related notations. The
equivalent width \ewint\ of the \emph{intrinsically} emitted spectrum is given
by the stellar population; for instance, a more top-heavy
IMF emits a harder spectrum, with relatively more ionizing and hence \lya\
photons. When escaping the galaxy, the spectrum has an equivalent width \ewem;
as we shall see below, the ISM may possess the ability of transferring an
uneven fraction of \lya\ vs.\ continuum photons, making \ewem\ larger or smaller
than \ewint. Finally, as the total number of photons in a line is conserved
when traveling through the Universe, whereas the number of continuum photons
per wavelength bin is reduced as they are cosmologically redshifted, the
\emph{observed} equivalent width \ewobs\ from a source at redshift $z$ is given
by $\ewobs = (1+z)\ewem$.


\subsection{The Neufeld scenario}
\label{sec:neufeld}

As \lya\ photons scatter on neutral hydrogen, their path length before escaping
a galaxy will always be longer than that of continuum radiation. Thus, the
immediate corollary is that \lya\ will be more susceptible to dust absorption
than the continuum, implying that $\ewem < \ewint$. However,
\citet[][hereafter N91]{neu91} investigated analytically the resonant scattering of \lya\
photons through the ISM and found that under special circumstances, the
\lya\ photon may actually suffer \emph{less} attenuation than radiation
which is not resonantly scattered, e.g.\ continuum radiation. \neu\ considered
the escape of radiation from the center of a plane-parallel, two-phase
structure, in which sperical clouds of homogeneously mixed neutral hydrogen and
dust lie embedded within a virtually empty intercloud medium (ICM). The number
density of clouds are assumed to be small enough that they do not touch, yet
sufficiently numerous that they cover most of the sky.
%

Furthermore, the source of both \lya\ and continuum photons is assumed to be
pointlike and situated in the center of the slab, in the ICM. Under these
conditions \neu\
showed that \lya\ photons, upon entering a cloud, scatters only a few times
before returning to the ICM, thereby being exposed arbitrarily little to
absorption by dust. Consequently, the journey of the \lya\ photons will be
confined primarily to the dustless ICM, preserving the total \lya\ luminosity.
In contrast, the continuum radiation which penetrates the clouds rather than
scattering off of their surfaces will be subject to the full attenuation, given
by the optical depth of dust from the center and out. 

Thus, since the continuum is reduced while the line is more or less preserved,
the EW of the escaping radiation can be ``boosted'' to arbitrarily high values.
We define such a boost as
\begin{equation}
\label{eq:b}
b \equiv \frac{\ewem}{\ewint}.
\end{equation}
Of course, the boost does not imply an increased \lya\ luminosity, but rather a
reduced continuum luminosity.


\subsection{Numerical and observational support}
\label{sec:numobs}

The mere potential of an EW boost was the conclusion of \neu's studies. 
\citet[][hereafter HO06]{han06} studied the scenario in more detail, both
analytically and numerically. In addition to confirming the boost, they also
looked into the effect of galactic outflows of clumpy gas on the line profile,
exploring the RT in various geometries.

While \han\ did investigate various extensions of the Neufeld scenario, one of
their main conclusion was the same as \neu: ``If most of the dust resides in a
neutral phase which is optically thick to \lya, the \lya\ EW can be
strongly enhanced''. The results of \neu\ and \han\ seem a natural explanation
of the unusually high observed EWs, and lately such observations has routinely
been interpreted within the framework of this scenario, inferring the presence,
or non-presence, of a clumpy, dusty ISM.
Thus, \citet{cha05} and lately \citet{bri12}, interpret the fact that \lya\
radiation is visible from $\sim 50$\% of a sample of sub-mm galaxies as
suggesting the presence of very patchy and inhomogeneous dust distribution.
\citet{fin07} regards a clumpy ISM as a possible explanation of their high
\lya\ EWs, and in \citet{fin08} propose a parameter $q$ to characterize
clumpiness, such that \lya\ is attenuated by a factor $e^{-q\tau}$, where
$\tau$ is the dust optical depth for non-resonant photons close to the \lya\
line. For a given \lya\ EW observation they then calculate the clumpiness of
the ISM, positing that for $q < 1$ the EW is boosted due to dust clumpiness.
This method is further used in \citet{fin09a,fin09c,fin11a}, \citet{yum10},
\citet{bla11}, \citet{nak12} and \citet{has12} to infer various degrees of
clumpiness in observed LAEs.
A equivalent approach was followed in \citet{nii09} and \citet{kob10}.
Similarly,
\citet{day08,day09,day10,day11} has invoked a clumpy ISM scenario in
order to match their simulated LAE LFs to observed ones.


However, the idealized conditions under which
a boost may be achieved is rarely taken into account. Accordingly, we
believe that a revision of the scenario is timely, and in the present work we
aim to investigate the validity of the various premises of the model. We do
this by calculating numerically the RT of \lya\ and continuum in model galaxies
constructed to cover a broad range of plausible as well as implausible systems.
In addition to the work of \neu\ and \han, the topic of resonant scattering in
a multiphase medium has received attention from \citet{ric03}, \citet{sur12},
\citet{dij12}, and \citet{duv12}.
Of these, only \citet{duv12} concern themselves with the relative escape
fraction of \lya\ and continuum, and found tentative evidence for the
impracticability of enhancing the \lya\ EW, by investigating the \lya\ and
continuum RT in an expanding shell of clumps with different values of dust
optical depth, expansion velocity, \HI\ density, and line width. They found
that indeed a virtually empty ICM was needed, along with only modest outflow
velocities and high \ebv, for an ISM geometry to result in a boost.

After describing the principles of the applied numerical code in \sec{RT},
we systematically vary all parameters relevant for decribing a (model) galaxy
in \sec{crit}, one at a time. This gives us a feeling for the impact a given
parameter has on the boost. However, as the change of one parameter may very
well be either enhanced or counteracted by the change of another, after
discussing in \sec{real} observational and theoretical constraints on the
actual values of the parameters that are likely to be met in real astrophysical
situations, we subsequently, in \sec{res}, undertake a large sample of
simulations with random parameters covering these values, constructing a
likelihood map of achievable boosts. The results of these calculations are
discussed in \sec{disc}, along with a discussion on other possible scenarios
that could lead to high \lya\ EWs.  


\section{Radiative transfer simulations}
\label{sec:RT}

\neu\ regarded each dusty cloud as a scattering particle itself, with a certain
probability of absorbing a photon. When \han\ developed their numerical model,
they followed the same approach. They used a Monte Carlo code, where the path
of individual photons were traced as they traveled through the ICM, but
whenever a cloud was encountered, the photon would simply be absorbed or
scattered at once in some direction away from the cloud. The appropriate
probability density functions (PDFs) were calculated on beforehand from fits to
a series of RT calculations of photons incident on a cloud surface. This
rendered feasible simulations that were otherwise impractical. In the almost
seven years
that have passed since, computational power has increased to a point that
``brute force'' simulations can in fact be performed rather effortlessly, 
obliviating the need for several approximations. 

\begin{figure}[!t]
\centering
\hspace{-5mm}
\includegraphics[width=0.26\textwidth]{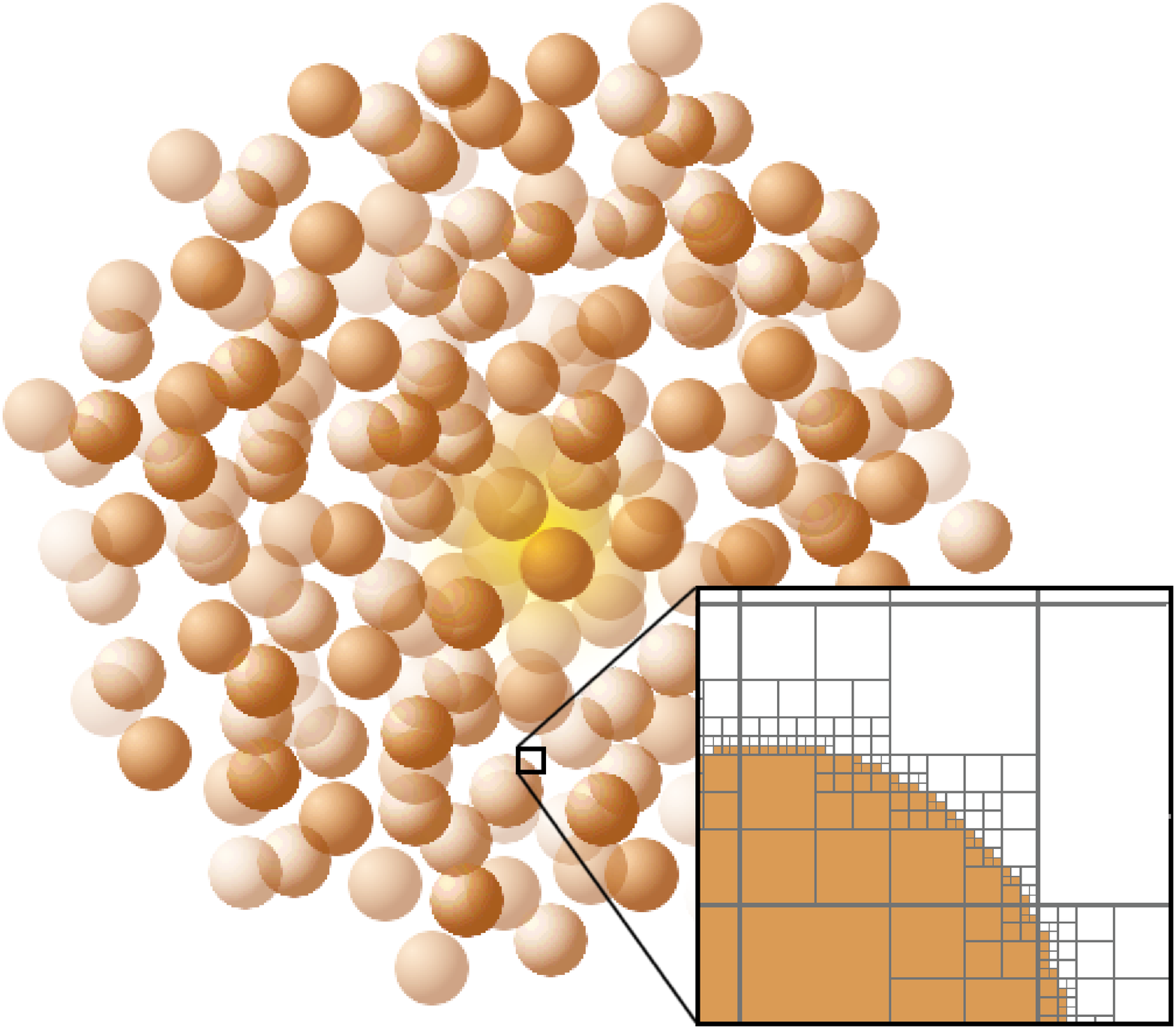}
\reflectbox{\includegraphics[width=0.23\textwidth, bb= 365 274 563 472]{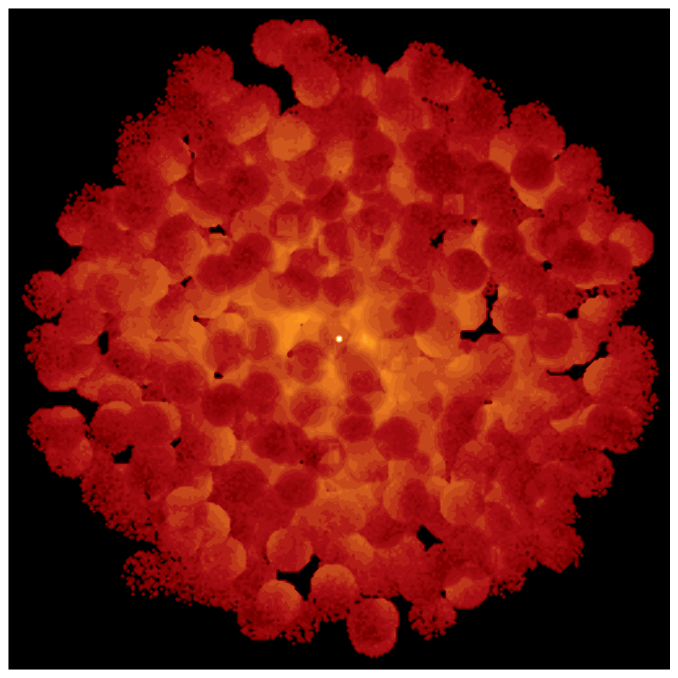}}
\caption{\emph{Left:} Graphical representation of the design of the model
  galaxy with $\sim 200$ clouds. The AMR structure with six levels of
  refinement is illustrated in the inset. Note that for presentation purposes
  only, a two-dimensional analogy is shown; in reality all simulations are
  carried out in three dimensions.
  \emph{Right:} Simulated surface brightness map of a model with $\sim 500$
  rather large clouds of radius $\rcl = 350$ pc to emphasize the
  structure. The fiducial model used in \sec{crit} has 6500 clouds of $\rcl =
  100$ pc, with a maximum refinement level of $\simeq 10$.}
\label{fig:AMR}
\end{figure}

In the present work, the RT calculations are conducted using the numerical code
\moca\ \citep{lau09a,lau09b}. In the following, the basics of the code are
outlined: 

The galaxy is constructed on a grid of cells, each of which holds a value of
the physical parameters important to the RT; the neutral hydrogen density \nHI,
the dust density \nd, the gas temperature $T$, and the three-dimensional
velocity $\mathbf{v}_{\mathrm{bulk}}$ of the gas elements. The original
Neufeld scenario concerned itself with radiation escaping from a slab of gas.
For numerical reasons, higher resolution can be achieved considering radiation
escaping a \emph{sphere} of multiphase gas, as did also \han. Qualitatively,
and even quantitatively except for factors of order unity, the results are
equivalent.  Thus, a galaxy is modeled as a number $N_{\mathrm{cl}}$ of
spherical, non-overlapping clouds with radius \rcl, dispersed randomly within a
sphere of radius \rgal.  A cell may be either an ICM cell or a cloud cell. In
order to make the clouds spherical, cells on the border between a cloud and the
ICM are adaptively refined, such that a given cell is split into eight
subcells, recursively until a satisfactory resolution is achieved. The
structure is depicted in \fig{AMR}, where also a surface brightness map of a
simulation is shown.
We typically consider $\sim 10^{3\textrm{-}5}$ clouds, with each cloud
typically consisting of $10^{2\textrm{-}3}$  cells. Note however that the
actual shape of a cloud does not affect the results significantly; the most
important quantities, as identified by \han, are the cloud albedo, i.e.\ the
probability that a photon incident on a cloud is reflected rather than absorbed
after a number of scatterings, and the average number $N_0$ of clouds with
which a \lya\ photon interacts before escaping the galaxy, in the absence of
absorption. This number in turn is a function of the covering factor \fc,
which is the average number of clouds intercepted by a sightline from the
center and out. Individual \lya\ and continuum photons are then emitted and
traced as they scatter in real and frequency space throughout the inhomogeneous
ISM.

\moca\ has previously been applied to galaxies extracted from
cosmological simulations. Although it was tested thoroughly against various
analytical solutions in \citet{lau09a,lau09b}, some modifications and
extensions had to be made in order to use it for the idealized simulations in
the present study. These are described and tested in App.~\ref{sec:moca}.

For a given set of parameters $N_{\mathrm{cl}}$, \rcl, etc., the
\emph{observed} boost will depend on the direction which it is observed, as
well as on the actual random realization of cloud positions. In addition to
sampling the average spectrum, i.e.\ the spectrum of photons escaping in all
directions, \moca\ calculates the spectra escaping exactly in the six different
directions along the Cartesian axes.
The more clouds a galaxy comprises, the smaller the spread will be in different
directions, and for different realizations.
Further, the variation between different directions is much larger than between
different realizations. For instance, for a covering factor of $\fc = 2$, only
$\sim 30$\% of the sightlines will actually intercept two clouds, while $\sim
10$\% will intercept no clouds at all (i.e.\ the covering \emph{fraction} is
$\sim 0.9$ for $\fc\sim2$), and a non-negligible fraction of the sightlines
will intercept $\geq 5$ clouds. These three cases will, respectively, result
in a boost close to the $4\pi$ average, no boost at all, and an essentially
infinite boost (for a central point source).
Thus, the fractional standard deviation $\sigma_b / b$ from different viewing
angles is typically of order 50\%, while for the $4\pi$ average of many
realizations of the same model, $\sigma_b / b$ is but a few percent.
In the following analysis, the presented values of $b$ represent the $4\pi$
average for a single realization of a given model.

\section{Investigating the criteria for a boost}
\label{sec:crit}


In order to systematize which physical conditions are necessary for boosting
the EW, we first settle on a fiducial model. In the subsequent sections, several of the implied parameters
are then relaxed or varied, individually or in conjunction. When nothing else
is stated, the remaining parameters correspond to the fiducial model.
Initially, we will not concern ourselves with the realism of the parameters,
instead deferring this discussion to \sec{real}, although we note that several
of the chosen values, as well as various derived parameters such as color
excess and total galaxy mass, roughly resemble observed typical LAE values.

The fiducial model galaxy is a sphere of radius $\rgal = 5$ kpc
consisting of $N_{\mathrm{cl}} = 6500$ of clouds of equal radii
$\rcl = 0.1$ kpc. The density of neutral hydrogen in the clouds is $\nHI = 1$
\pcmc, implying a column density $\NHIcl\sim3\times10^{20}$ \pcms\ as measured
from the center to the surface of a cloud.
The clouds are dispersed at
random in an empty ICM. The temperatures of the clouds and the ICM are
$T_{\mathrm{cl}} = 10^4$ and $T_{\mathrm{ICM}} = 10^6$ K, respectively.
The physical significance of the clouds and the ICM are the interstellar phases
conventionally called the warm neutral medium (WNM) and the hot ionized medium
(HIM), first identified by \citet{fie69} and \citet{mck77}, although in reality
\nHIcl\ is usually somewhat smaller than the chosen 1 atom \pcmc.
Finally, in the fiducial model, all
photons are emitted in the line center, from the ICM in the center of the
sphere.

The covering factor \fc\ is analogous to an optical depth of clouds intercepted
by a sightline from the center of the galaxy and out. With a number density
$n_{\mathrm{cl}} = N_{\mathrm{cl}}/V_{\mathrm{gal}}$ of clouds, the covering
factor of the fiducial model is then
$\fc = n_{\mathrm{cl}}\rgal\sigma_{\mathrm{cl}}
     = \frac{3}{4}N_{\mathrm{cl}}(\rcl/\rgal)^2 \simeq 2$
(where $\sigma_{\mathrm{cl}}$ is the cross section of a cloud).
Since an average sightline passing through a cloud traverses a distance
$\ave{d} = V_{\mathrm{cl}}/\sigma_{\mathrm{cl}} = 4\rcl/3$ inside the cloud,
the total path traveled inside clouds for an average sightline is
$\ave{d_{\mathrm{in\ clouds}}} = \fc\ave{d} = N_{\mathrm{cl}} \rcl^3 / \rgal^2$.

At the heart of the EW boosting mechanism lies the assumption of the \lya\
photons being shielded from the dust by neutral hydrogen while the
non-scattered FUV continuum radiation is subject to the ``full'' dust
absorption. Hence, the relevant quantity for dust absorption is the total
average absorption optical depth \taua\ of the dust throughout the galaxy.
The dust absorption optical depth of a single cloud (center-to-face) is
$\tau_{\mathrm{a,cl}} = (1-A) \tau_{\mathrm{d,cl}}$, where $\tau_{\mathrm{d,cl}}$ is the
total (absorption + scattering) dust optical depth, and $A$ is the dust albedo.
Hence, $\taua \simeq \tau_{\mathrm{a,cl}}\ave{d_{\mathrm{in\ clouds}}}/\rcl$
(this expression is only approximate, since photons scattering on dust grains
may still be subject to absorption by another dust grain before it escapes the
galaxy).

Dust grains are built from metals, and the density of dust is thus assumed to
scale with metallicity and hydrogen density.
Since observationally the optical depth is not readily measured, we will
instead refer to the metallicity $Z$, which is usually more easily probed.
Depending on the actual
extinction curve used, the constant of proportionality will differ.
Furthermore, assuming that the density of dust scales linearly with the
metallicity, the dust optical depth can be related to $Z$ through
$\tau_{\mathrm{d}} = N_{\mathrm{H}}\,\sigma_{\mathrm{d}}(\lambda)\,Z/Z_0$, where
$\sigma_{\mathrm{d}}$ is the dust cross section \emph{per hydrogen nucleus}
--- specific for a given extinction curve --- and $Z_0$ is the reference
metallicity of that extinction curve. For example, for SMC dust, at the \lya\
wavelength the cross section is $\sigma_{\mathrm{d}} = 4\times10^{-22}$ \cm2
with only a small wavelength depence across the line, and the SMC metallicity
is $Z_0 \simeq 0.25 \Zsun$.

Having explicated the basics of the model, we now proceed to investigate the
impact on the boost of varying the implied parameters.

\subsection{Cloud covering factor}
\label{sec:fc_invest}

\begin{figure}[!t]
\centering
\includegraphics [width=0.40\textwidth] {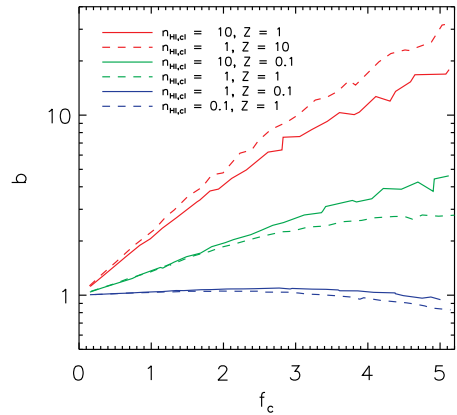}
\caption{EW boost $b$ as a function of covering factor $f_{\mathrm{c}}$, for
  various values of cloud \HI\ densities \nHIcl\ and metallicites
  $Z$. Lines with the same color share the same value of dust optical depth
  per cloud: \emph{blue}, \emph{green}, and \emph{red} denote
  $\tau_{\mathrm{a,cl}} \simeq 0.02$, 0.2, and 2, respectively. For these
  simulations, the number of clouds and the galaxy radius was $N_{\mathrm{cl}}
  = 100$--50\,000 and $\rgal = 2$--10 kpc. Densities are given in \pcmc, and
  metallicites in terms of \Zsun.}
\label{fig:fc}
\end{figure}

For the Neufeld scenario to be efficient, the sky, as seen from the
center of the galaxy, must be sufficiently covered by clouds. \neu\ asserts that
the covering factor \fc\ must be larger than unity. In fact, as soon as a
single cloud is present, the $4\pi$ average will have $b > 1$. \Fig{fc} shows,
for various values of \nHIcl\ and $Z$ how the EW boost changes with \fc.
The boosts are seen to follow an approximate log-normal relation, where
different combinations of \nHIcl\ and $Z$ with equal dust optical depths lie
roughly on the same line.
At fixed $Z$, larger \HI\ densities result in larger boosts. The \lya\
photons are exposed to the same amount of extinction, but since \taua\ is
larger, the continuum is reduced more.
At fixed \nHIcl, a larger $Z$ also results in a larger boost due to the higher
\taua, but since for each cloud interaction a \lya\ experiences a larger
probability of being absorbed, combinations with high $Z$ have smaller boosts
than those with high \nHIcl. This is especially true at large values of \fc,
where the \lya\ photons interact with many clouds.


\subsection{Cloud hydrogen density}
\label{sec:nHIcl_invest}

To shield the \lya\ photon from the dust, the clouds must be highly optically
thick in hydrogen. Since non-absorbed photons are usually reflected after only
a handful of scatterings, in principle the clouds could have an optical depth
of only a few, although in that case, for realistic metallicites, the dust
optical depth would be so low that the continuum passes unhindered
through the clouds.

\begin{figure}[!t]
\centering
\includegraphics [width=0.40\textwidth] {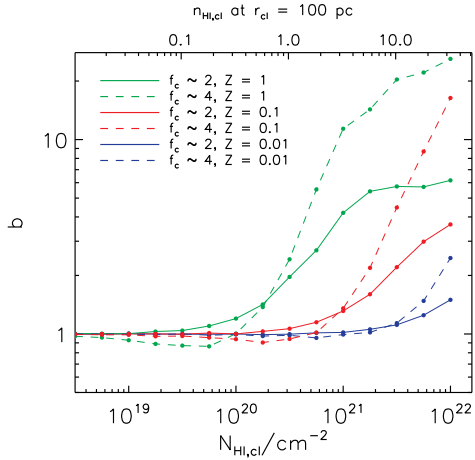}
\caption{EW boost $b$ as a function of neutral hydrogen column density \NHIcl\
  of the individual clouds, as measured from the center to the surface of a
  cloud, for different gas metallicites. The upper $x$ axis gives the
  corresponding number densities in \pcmc\ in the case of $\rcl=100$ pc.
  Metallicities are given in terms of \Zsun. The simulations for which $\fc
  \sim 4$ have been realized simply using $N_{\mathrm{cl}} = 13\,000$.}
\label{fig:NHI_cl}
\end{figure}

\Fig{NHI_cl} shows how the magnitude of the boost depends on the \HI\ (column)
density of the clouds. Note that for a given series of simulations, the
metallicity has been held fixed, meaning that \taua\ increases with increasing
\NHIcl.
Even for Solar metallicity, the EW boost is seen to set in only after \NHIcl\
surpasses approximately $10^{20}$ \pcms, while, say, ten times lower
metallicity requires ten times higher \NHIcl. The critical value for a
noticeable boost is found where \NHIcl\ and $Z$ conspire to a dust optical
depth of order unity.  For very high values of \NHIcl\ the clouds become so
optically thick in dust to continuum such that the calculated EWs become rather
noisy.


\subsection{Dust contents in the clouds}
\label{sec:Zcl_invest}

\begin{figure}[!t]
\centering
\includegraphics [width=0.40\textwidth] {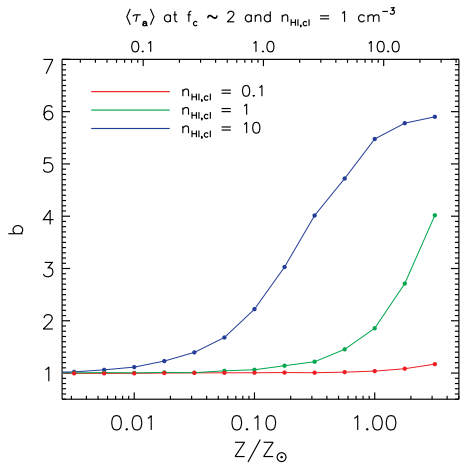}
\caption{EW boost $b$ as a function of gas metallicity $Z$ --- which can be
  taken as a proxy for \taua\ --- for different \HI\ densities \nHIcl\ of the
  clouds.}
\label{fig:Z}
\end{figure}

The density of dust scales both with metallicity and \HI\
density. Increasing the gas density in the clouds will increase the dust
density accordingly. This does not change the surface of a cloud from the point
of view of a \lya\ photon, but the continuum photons experiences a larger total
optical depth of dust, and thus the boost increases. For a given \HI\ density,
increasing the metallicity results in a higher boost. These relations are shown
in \fig{Z}.
As expected, larger \HI\ densities result in larger boosts, but for
sufficiently low metallicites the boost will disappear. Again, the critical
value of $Z$ is the metallicity required for the optical depth of dust to be of
order unity. For the adopted cloud size of 100 pc, this corresponds to the
product $\nHI\,Z/\Zsun \sim 1$, which is indeed seen in \fig{Z}.


\subsection{Cloud size distribution}
\label{sec:rcl_invest}

\begin{figure}[!t]
\centering
\includegraphics [width=0.40\textwidth] {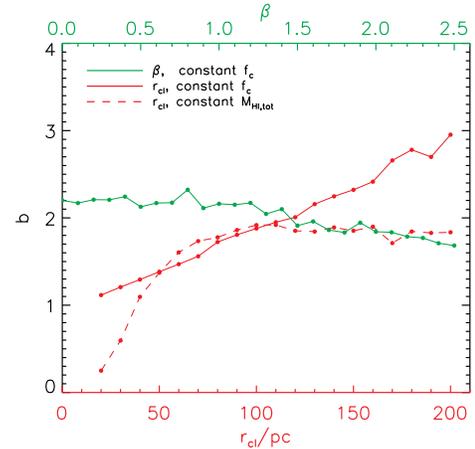}
\caption{\emph{Lower $x$ axis and red lines}: EW boost $b$ as a function of
  cloud radius \rcl, maintaining constant covering factor \fc\ (\emph{solid})
  and a constant total \ion{H}{i} mass (\emph{dashed}).
  \emph{Upper $x$ axis and green line}: EW boost as a function of cloud size
  distribution power-law
  index $\beta$, for minimum and maximum cloud radii $r_{\mathrm{min}} = 20$ pc
  and $r_{\mathrm{max}} = 200$ pc, at constant \fc.}
\label{fig:beta_and_rcl}
\end{figure}

In the previous simulations the cloud radii were held constant at $\rcl = 100$
pc.
As \neu\ proposed and \han\ confirmed through their numerical model, the
covering factor rather than the actual shape and size of the clouds is the
important parameter controlling the RT of the \lya\ photons. For the continuum,
however, doubling the cloud radius at constant \fc\ doubles the dust optical
depth, thus increasing the boost.
In \fig{beta_and_rcl} we
investigate the dependency of the boost on cloud \emph{sizes}, as well as on
cloud size \emph{distributions} $P(\rcl) \propto \rcl^{-\beta}$.
The dependency on slope $\beta$ is seen to be small. In the simulations with
varying cloud radii, the slope was held constant at $\beta = 2$. This value was
chosen to follow \han.

Increasing \rcl\ or decreasing $\beta$ at constant \fc\ both corresponds to in
increasing the \taud\ and hence $b$, although it is seen that changes in
$\beta$ affects $b$ only minutely. On the other hand, changing the cloud sizes
but maintaining a constant total \ion{H}{i} mass implies a constant \taud, and
hence a constant continuum escape fraction. Consequently, the boost does not
change for $\rcl \gtrsim 100$ pc, but at lower \rcl\ the number of times that
the \lya\ photon interact with clouds ($N_0$) becomes so large that the
\lya\ escape fraction decreases considerably, reducing the boost.

%


\subsection{Cloud velocity dispersion}
\label{sec:sigVcl_invest}


The very concept of the boost hinges on the fact that neutral gas shields
photons close to the line center from the dust. But if a given cloud as a whole
has a non-vanishing bulk velocity, the entire spectrum is Doppler shifted in the
reference frame of the cloud, and for sufficiently large velocities, all line
photons suddenly are no longer in resonance. In general, \HI\ regions may
be expected to exhibit such macroscopic motions, described by a cloud velocity
dispersion \sigV.

\begin{figure}[!t]
\centering
\includegraphics [width=0.40\textwidth] {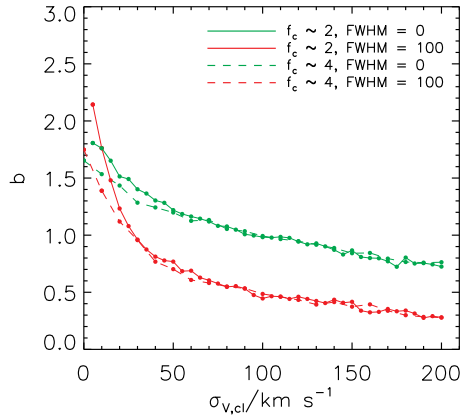}
\caption{EW boost as a function of cloud velocity dispersion \sigV, for
  zero-width (\emph{solid}) and broadened (\emph{dashed}) emission, in
  galaxies of $\fc\sim2$ (\emph{green}) and 4 (\emph{red}).}
\label{fig:sigV}
\end{figure}

\Fig{sigV} demonstrates that this is indeed the case; for $\sigV \gtrsim 100$
\kms, the fiducial model is seen to be incapable of boosting the EW. Moreover,
whereas in the case of $\sigV = 0$ a higher covering factor results
in a larger boost, introducing random cloud motion reduces the boost even
faster; for \fc\ twice that of the fiducial model, the boost vanishes already
for $\sigV \gtrsim 30$ \kms.
%


\subsection{Cloud gas temperature}
\label{sec:Tcl_invest}

\begin{figure}[!t]
\centering
\includegraphics [width=0.40\textwidth] {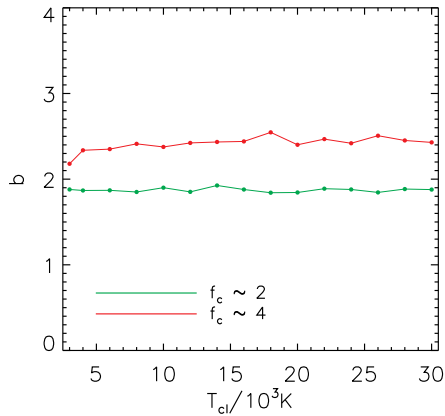}
\caption{EW boost as a function of cloud gas temperature $T_{\mathrm{cl}}$
  (including also small scale turbulence) for a covering factor $\fc\sim2$
  (\emph{green}) and $\fc\sim4$ (\emph{red}).}
\label{fig:T_cl}
\end{figure}

Increasing the gas temperature\footnote{The parameter ``temperature'' covers
also sub-grid turbulent motion, since this effectively broadens the line
profile by a Gaussian.} $T_{\mathrm{cl}}$ in the clouds means less atoms with
the right velocity for scattering photons exactly at the line center, in return
for more atoms available for scattering off center photons. Photons in the
wings of the line, however, do not care about the gas temperature, since the
profile here is given by natural broadening. Even so, the difference in the
effective cross section of the atoms is not very large. The consequence is that,
for increasing $T_{\mathrm{cl}}$, the photons penetrate slightly deeper into
the cloud, but not enough to affect the boost, as seen in \fig{T_cl}.
The only difference is that the line profile of the escaping radation is
somewhat broadened.


\subsection{Galactic outflows and infall}
\label{sec:Vout_invest}

Gas elements also move in large-scale, collective motions, seen e.g.\ in
galactic outflows \citep{joh71,rup02,sha03,vei05}. For the same reason as in
the previous section, such motions should act such as to diminish the boost.
This stands in contrast to what is expected \citep{kun98,dij10} for a
more homogeneous shell of gas being expelled from the galaxy, surrounding a
central source and covering the full sky. In that case, the \lya\ photons that
would otherwise have to scatter their way through the shell, being very
vulnerable to dust absorption, may be shifted away from the line center and
escape through the shell with minimal absorption.

\begin{figure}[!t]
\centering
\includegraphics [width=0.40\textwidth] {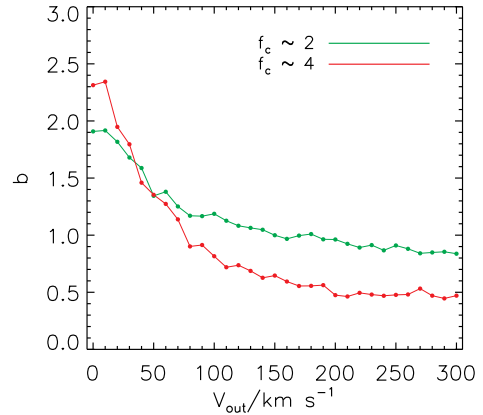}
\caption{EW boost as a function of terminal outflow expansion velocity
  \Vout for covering factors $\fc\sim2$ (\emph{green}) and $\sim4$
  \emph{red}.}
\label{fig:wind}
\end{figure}

\Fig{wind} shows how the EW boost depends on the expansion velocity \Vout.
The wind speed is a function of distance $r$ from the
center and is modelled assumed that the gas elements receive an acceleration $a
\propto r^{-\alpha}$ \citep{ste10,dij12}. The speed thus increases from 0 at
$r=0$ to the terminal velocity \Vout at $r=\rgal$. Note, however,
that the exact wind profile is not a major determinant of the resulting boost.

The curves in \fig{wind} resemble those in \fig{sigV}, although they are
somewhat more shallow. The reason is that a photon bouncing off of a cloud in
a galactic wind will, in general, upon its next encounter with a cloud, meet a
cloud with roughly the same velocity. In the case of random motions,
the relative velocity of the next cloud may be very high, increasing the
probability of the photon being absorbed.

For reasons of symmetry, in terms of $b$ the same results are obtained if the
velocities are inverted, as would be the case for galactic gas accretion
\citep[e.g.][]{dij06b}. The spectra, however, would be reflected about the line
center.


\subsection{ICM hydrogen density}
\label{sec:nHIICM_invest}

Thus far, we have assumed an ICM completely void of both dust and neutral gas.
Due to the high temperature of the ICM, the hydrogen is expected to be quite
highly ionized. Similarly, due to the high temperature, as well as the stronger
ionizing UV radiation field, dust may be expected to be partly depleted.
However, even a single scattering in the ICM may be fatal to a \lya\ photon:
Exactly because of the high temperature, a scattering event in the ICM is
likely to occur on a high-velocity atom moving more or less perpendicular to
the path of the photon, such that the velocity relative to the trajectory of
the photon is small. Unless the photon is scattered close to (or opposite
to) the same direction, it will be highly Doppler-shifted, such that next time
it encounters a cloud, it will penetrate deeply into the dusty medium, with a
higher probability of being absorbed. This effect is investigated further in
\sec{TICM_invest}.

\begin{figure}[!t]
\centering
\includegraphics [width=0.40\textwidth] {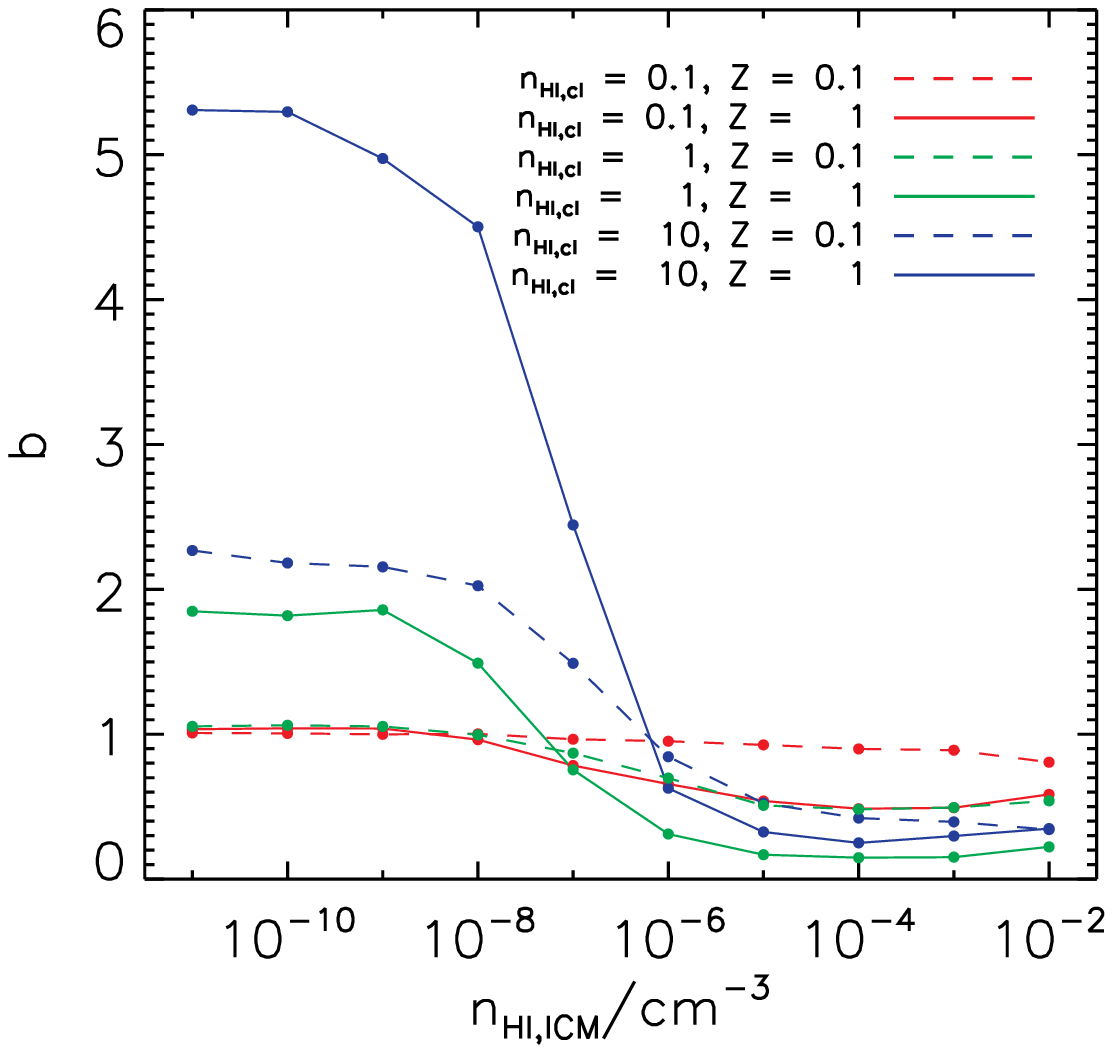}
\caption{EW boost $b$ as a function of \HI\ density \NHIICM\ in the ICM.
  Results are show in \emph{red}, \emph{green}, and \emph{blue}
  for cloud number densities $\nHIcl = 0.1$, 1, and 10 \pcmc, respectively, in
  combination with cloud metallicites $Z = 0.1$ \Zsun (\emph{dashed}) and 1
  \Zsun\ (\emph{solid}).}
\label{fig:NHI_ICM}
\end{figure}

For our fiducial model's radius of $\rgal = 5$ kpc, an optical depth of order
unity in the ICM for a line center photon is reached for an \HI\ density $\nHI
\sim 10^{-8}$ \pcmc.  Due to the resonance nature of the scattering, somewhat
higher densities are allowed as soon as the photon has diffused a few Doppler
width from the line center. This effect is visible if \fig{NHI_ICM}, where the
boost is shown as a function of ICM \HI\ density for different cloud densities
and metallicites.
%


\subsection{ICM dust density}
\label{sec:ZICM_invest}

The neglect of dust in the ICM may be justified by the fact that dust grains
will tend to get destroyed in hot and ionized media. Nevertheless, dust does
exist in \HII\ regions, albeit generally at lower densities. For an
increasing covering factor the photons' paths through the ICM is
increased as they walk randomly out through the galaxy. The mean free path
between each cloud interaction is $1/\fc$, while the average number of cloud
interactions is $N_0 = \fc^2 + \frac{4}{5}\fc$ (see App.~\ref{sec:test}). Thus,
the average path length for a photon out of our fiducial model of $\rgal=5$ kpc
is $\dtot \simeq 14$ kpc. For a density of 0.01 \pcmc, $Z = \Zsun$, and no dust
destruction, the total optical depth of dust actually reaches
$\ave{\tau_{\mathrm{ICM}}} \sim 0.2$. For larger densities \rgal\ and \fc,
$\ave{\tau_{\mathrm{ICM}}}$ could exceed unity, while if part of the dust is
destroyed, it may be neglected altogether.

\begin{figure}[!t]
\centering
\includegraphics [width=0.40\textwidth] {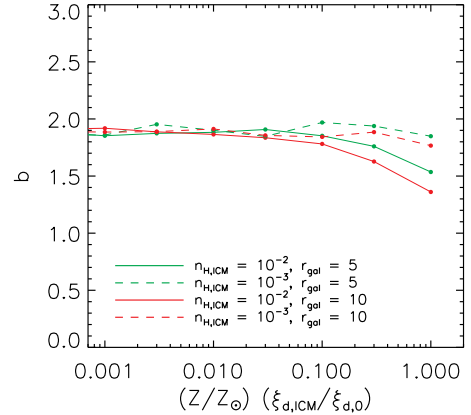}
\caption{EW boost $b$ as a function of dust density in the intercloud medium,
  parametrized through the amount of metals in the form of dust, $(Z/\Zsun)
  \times (\xidICM/\xi_{\mathrm{d,0}})$, where the last term is the
  dust-to-metal ratio in terms of the values typically found in the
  predominantly neutral environments from where the dust extinction curve is
  taken (here the SMC).  Results are shown for galaxies of $\rgal = 5$
  (\emph{green}) and 10 kpc (\emph{red}), in combination with total (neutral +
  ionized) hydrogen density $\nHICM = 10^{-3}$ (\emph{dashed}) and $\nHICM
  = 10^{-2}$ \pcmc\ (\emph{solid}).}
\label{fig:ZICM}
\end{figure}

\Fig{ZICM} shows the effect of dust in the ICM for various densities, galaxy
radii, and dust depletion factors. The important quantity is the number density
of dust, which is proportional to the product of total density, metallicity,
and dust-to-metal ratio \xid; that is, a given dust density can be realized in
several ways. While in predominantly neutral regions, \xid\ is rather constant
(see discussion in \sec{Zcl_real} and \sec{ZICM_real}). Accordingly, the boost
is shown as a function of the product of $Z$ and \xid, for different total
hydrogen densities.
Indeed, ICM dust is seen to be important only for rather high ICM densities,
large galaxies, and low dust depletion factors.
Thus we see that the most crucial factor
of the Neufeld mechanism is not, as it is often laid out, that the \lya\
photons are confined to a dustless medium but rather that are confined to a
medium of low density but possibly with the same dust-to-gas ratio, and high
ionization fraction.


\subsection{ICM gas temperature}
\label{sec:TICM_invest}

\begin{figure}[!t]
\centering
\includegraphics [width=0.40\textwidth] {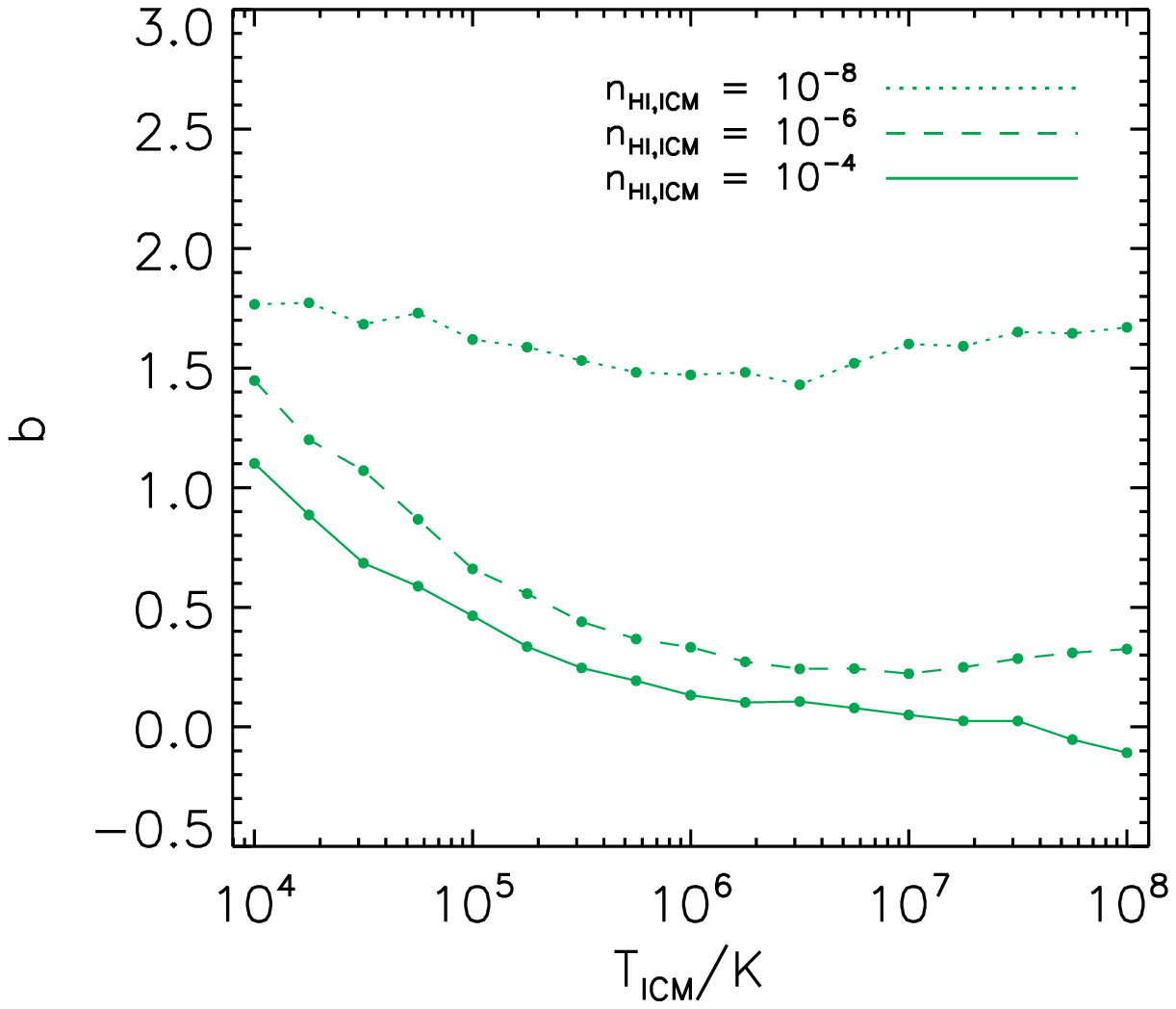}
\caption{EW boost as a function of temperature $T_{\mathrm{ICM}}$ in the ICM,
  for neutral hydrogen densities $\nHIICM = 10^{-8}$ (\emph{dotted}), $\nHIICM
  = 10^{-6}$ (\emph{dashed}), and $\nHIICM = 10^{-4}$ (\emph{solid}) \pcmc.}
\label{fig:T_ICM}
\end{figure}

As mentioned in \sec{nHIICM_invest}, although scatterings in the ICM evidently
are rare a single such event may easily result in the subsequent absorption of
the photon. \Fig{T_ICM} explores this effect, but it is seen that only for
rather high values of \nHIICM, the boost decreases significantly.
For low densities, scatterings are so rare that even though some photons are
absorbed, it does not affect the net result much. For very high temperatures,
the boost increases again, since the number of atoms available with the right
velocity to scatter a photon becomes too small.


\subsection{Location of emission}
\label{sec:Xinit_invest}

\begin{figure}[!t]
\centering
\includegraphics [width=0.40\textwidth] {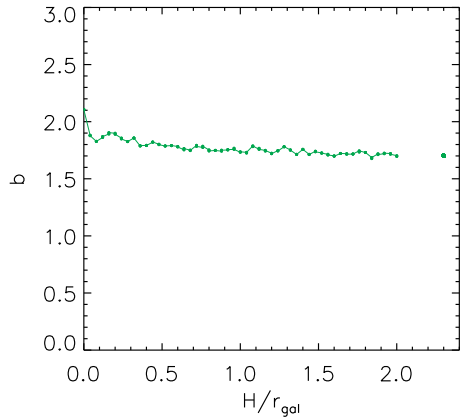}
\caption{The effect of an extended emission, as given by an exponentially
  decreasing emission site PDF of the form $P(r) \propto e^{-r/H}$, where $r$
  is the distance from the center of the galaxy, and $H$ is the scale length.
  The rightmost point shows the boost for $H \rightarrow\infty$, i.e.\ for a
  completely homogeneous distribution of sources.}
\label{fig:grad}
\end{figure}

Although star formation tends to be centrally located \citep[e.g.][]{fru06},
obviously
the photons are not emitted from a point source in reality. However, as long as
the initial emission direction is isotropically distributed, photons emitted in
the outskirts of the galaxy may be emitted both in an outward direction, being
subject to a lower covering factor and thus a smaller boost, or toward the
center, resulting in a higher covering factor and higher boost. As the number
of cloud interactions scale non-linearly with \fc, however, an extended
emission profile does result in a slightly lower boost, although the effect is
miniscule, as seen in \fig{grad}.

\begin{figure}[!t]
\centering
\includegraphics [width=0.40\textwidth] {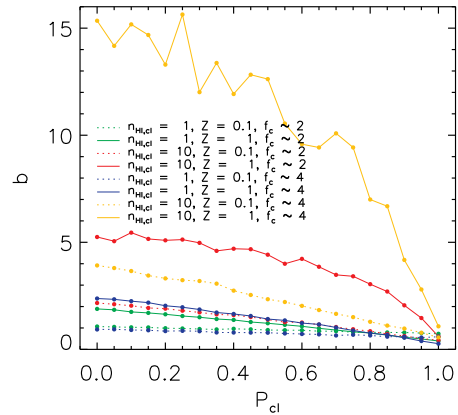}
\caption{EW boost $b$ as a function of probability \Pcl\ for a
  photon to be emitted from inside a cloud. An emission PDF with a scale length
  of $H = 1$~kpc has been adopted. For $\Pcl = 0$, the model reduces
  to the fiducial model, except for the somewhat extended emission which has
  only a modest influence on $b$ (cf.\ \fig{grad}). For $\Pcl = 1$, all photons
  are emitted from the interior of clouds. Results are show for all
  combinations of $\fc \sim 2, 4$, $\nHIcl/\pcmc = 1, 10$, and $Z/\Zsun = 0.1,
  1$.}
\label{fig:Pcl}
\end{figure}

While the extended emission relaxes the notion of a central point source, all
photons were still emitted from the ICM, as in the case of the central source
in all previous sections. However, as the \lya\ radiation is assumed to
originate from gas surrounding young stars, which form from gas clouds that
have recently cooled sufficiently to initiate star formation, it is
probably more realistic to have at least a fraction of the photons originating
from \emph{within} the clouds. In \fig{Pcl} we investigate how the EW boost
changes for an increasing correlation of the photon sources with the clouds.
As expected, taking this effect into account diminishes the boost, as the
path out of a cloud is longer for a \lya\ photon than for a continuum photon.


\subsection{Intrinsic line profile}
\label{sec:sigline_invest}

In the fiducial model, all \lya\ photons are emitted exactly in the line
center, i.e.\ as a delta function. In reality, the intrinsic line will have a
finite width, which is convolution between the natural, Lorentzian line profile
and the thermal, Gaussian distribution of the atom velocities. Moreover,
macroscopic motion of the gas parcels emitting the photon will cause an
effective broadening of the line. As the temperature of the emitting gas is of
the order $10^4$ K, corresponding to a line width of $\sim 10$ \kms, the
macroscopic motion may dominate over the thermal (see \sec{nHIcl_real}). From
non-resonant lines such as H$\alpha$, we know that the intrinsic line width may
easily reach tens of \kms, i.e.\ several Doppler widths. As photons born this
far from the line center are not efficiently shielded from the dust, they are
more prone to dust absorption.

\begin{figure}[!t]
\centering
\includegraphics [width=0.40\textwidth] {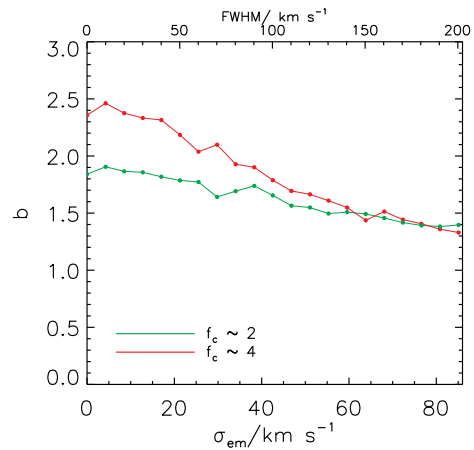}
\caption{EW boost $b$ as a function of line width $\sigma_{\mathrm{em}}$ for
  covering factors $\fc\simeq2$ (\emph{green}) and 4 (\emph{red}).}
\label{fig:fwhm}
\end{figure}

The effect of a broadened emission line profile is shown in \fig{fwhm}.
%



\section{Toward a realistic model}
\label{sec:real}

Whereas the fiducial model in the previous section was not chosen to be
particular realistic, but rather to roughly resemble the original Neufeld
scenario, we now proceed to investigate what values of the physical parameters
we in fact expect to meet in real LAEs. 
In building a ``realistic'' model, it should be kept in mind that several of
the input parameters are mutually dependent. For instance, under the assumption
of approximate pressure equilibrium, the temperatures and densities are related
as $n_{\mathrm{tot,cl}} T_{\mathrm{cl}} \sim n_{\mathrm{tot,ICM}}
T_{\mathrm{ICM}}$, where subscripts ``tot'' refer to the total number density
of all particles. The \HI\ densities are then given by the ionization fractions
which, in turn, depend on not only temperature, but also the chemical
composition, i.e.\ on metallicity.  Note, however, that due to very different
cooling time scales of the WNM and HIM, the different phases of the ISM need
not be in thermal equilibrium.  Furthermore, the dependencies of some of the
parameters may actually serve such as to counteract a boost.
For instance, a larger galaxy in general entail a larger \fc, which may
increase $b$, but since the larger mass also typically implies a larger
\sigV, $b$ is quickly reduced again.
Similarly, although a increased metallicity tends to increase $b$,
metals also donate free electrons, which aids to increase the neutral fraction
of hydrogen in the ICM, thus decreasing $b$.

The approach taken here is to identify for each parameter the minimum and the
maximum values conceived to be encountered in LAEs, as well as more typical
ranges. Subsequently a series of RT simulations is run with the physical
parameters taking random values in the chosen ranges, and finally regions of
high and low likelihood are identified in the parameter space, exploring
whether any of these models lead to a significant boost.



LAEs span a wide range of physical characteristics, and in many aspects it does
not make sense to speak about a ``typical LAE''. It is often stated that LAEs
are quite young, fairly small, relatively highly star-forming galaxies of
rather low metallicity and dust contents. However, one
must bear in mind that many of these aspects are a consequence of the selection
criterion picking out objects at high redshifts, i.e.\ at early epochs.
In the following sections, values and ranges of the relevant parameters argued
to be more or less representative of LAEs are summarized from the literature.
An overview of the values is found in \tab{params} on page
\pageref{tab:params}.

\subsection{Cloud hydrogen density and temperature}
\label{sec:nHIcl_real}

Gas does not cool equally well at all temperatures. Rather the cooling function
exhibit several plateaus at which the gas temperature tends to settle. This is
part of the reason for the different phases of the ISM; for gas which is in
rough pressure equilibrium, the densities of the phases are set by their
temperature, and thus, although variations do exist, they are not expected to
span many orders of magnitude within a given phase.
In the WNM, to which the clouds in the model correspond, characteristic
hydrogen densities are 0.2--0.5 \citep[e.g.][]{car98,fer01,glo04} --- typically
probed by \ion{H}{i} 21 cm emission --- with a neutral fraction close to unity.
At low (high) pressure, \nHI\ may reach values of $\sim0.02$ (3) \pcmc
\citep{wol03}. If densities rise much above $\sim1$ particle \pcmc, cooling
becomes so efficient that the gas will tend to contract further to a cold
neutral medium (CNM) and molecular clouds, thus quickly decreasing \fc.
The temperature of the WNM is of the order $10^4$ K, ranging
from 8\,000 K to 12\,000. Since small scale turbulence effectively broadens the
line in the same way as thermal motion, this process can be taken into account
using a higher temperature. Turbulence is usually of the order of (a few times)
the speed of sound \citep[e.g.][]{gae11} which, in turn, is the same order as
the thermal motion. The effect can thus be simulated by letting $T =
2$--$3\times10^4$ K. Note also that much of the WNM may be in a thermally
unstable phase \citep{hei03}, reaching temperatures below 5\,000 K.


\subsection{ICM hydrogen density and temperature}
\label{sec:nHIICM_real}

Surrounding the neutral gas is the HIM, heated and ionized to a large degree
by supernova shock waves sweeping through the ISM. Expanding due to the larger
pressure, and thus diluting, the cooling timescales are so long that
temperatures of $5\times10^5$--$5\times10^6$ K and higher is easily maintained
\citep{bri00,tul06,tul08}. Densities in the range $10^{-3}$ to $10^{-2}$ are
typical here \citep{dop03,fer01}, but since the fraction of neutral hydrogen
at these temperatures is of the order $\xHIICM\sim10^{-8}$--$10^{-5}$
\citep{hou64,sut93}, scattering in the ICM is very rare.
These properties are usually probed through X-ray emission
\citep[e.g.][]{tul06}, or absorption lines
of highly ionized metals \citep[e.g.][]{par09}.
Note that the shock waves push the phases out of
equilibrium and may in fact dominate the state of the ISM \citep{dea05}.


\subsection{Covering factor}
\label{sec:fc_real}

A typical covering factor is difficult to establish, but a hint is provided by
spectra of sources lying behind intervening galaxies.
For instance, \citet{not12} fit low-ionization absorption lines such as
\ion{Si}{ii} --- originating from cool clouds --- in a quasar spectrum, the
line of sight toward which passes through a DLA lying at $z=2.2$, with 5--8
components. The impact parameter of the sightline with respect to the central
emission of the DLA galaxy counterpart is pretty small ($\lesssim 1$ kpc), and
thus in this case the covering factor is roughly $\fc = 2.5$--4.

In the Milky Way (MW), the number of neutral clouds per kpc in the plane is
roughly 4--8 \citep{knu79,knu81,fra12}. Although this implies a high covering
factor along a sightline in the plane, perpendicular hereto the covering factor
is only of order unity. Since the disk has formed from a collapsed sphere, one
may expect a smaller number of clouds per kpc for a spherical system. For
instance, the scale height of the MW's WNM is approximately $h = 300$ pc and
its radius is $R_{\mathrm{MW}} \sim 15$ kpc \citep{bin98}; distributing the
clouds over a sphere of the same radius reduces the crowdedness of clouds by a
factor of $\sim R_{\mathrm{MW}} / h = 50$. 


\subsection{Cloud metallicity and dust density}
\label{sec:Zcl_real}

Generally lying at high redshifts where metals have had less time to build up,
LAEs tend to have lower metallicities, and hence dust densities, than local
galaxies. Measured values of $Z$ range from metallicities as low as $\sim
10^{-2} \Zsun$ to values similar to local values, and in some cases even
supersolar metallicities have been found at high redshifts
\citep[e.g.][]{acq12}. In a survey of $\sim10^3$ LAEs at $z = 2.2$,
\citet{nak12} found a lower limit for the average metallicity of a LAE of
$Z = 0.09 \Zsun$. Even low-redshift LAEs tend to have modest metallicities
\citep[$Z \simeq 0.1$--1][]{gia96,ost09}.

Correspondingly, measured color excesses tend to be modest. For example, at
$z \sim 3$ \citet{gro07} and \citet{pir07} found \ebv's lying approximately in
the range 0.01 to 0.1. \citet{ver08} found similar values by fitting \lya\ line
profiles using the \lya\ RT code {\sc MCLya} \citep{ver06}. At $z = 2.2$,
\citet{hay10} found values spanning all the way from 0 to $\sim 0.3$. For
comparison, our fiducial model has $\ebv = 0.05$.

In our model we assume that the dust-to-metal ratio \xidcl\ in the clouds is
equal to that of our reference extinction curve, such that $\tau_{\mathrm{d}}$
scales linearly with $Z$. Observations of \xidcl\ at high redshifts are sparse,
but tend to be similar \citep[e.g.][]{pet97,sav03} or slightly lower. Various
analytical and numerical calculations predict a metallicity-dependent evolution
of the dust-to-metal ratio, such that \xidcl\ reaches present-day values only
after timescales of 10--100 Myr \citep{gal11} or even several Gyr \citep[][see
also \citet{mat11}]{ino03}. Moreover, measurements in the local but very
low-metallicity galaxy I Zw 18 indicate that \xidcl\ is lower in such galaxies.
Since we have used the extinction law of the SMC, which is also a
low-metallicity galaxy with a rather young stellar population, we expect that
its extinction will not be substantially different from that of high-redshift
galaxies.  


\subsection{Cloud velocity dispersion}
\label{sec:sigVcl_real}

From the virial theorem, the components of a galaxy of mass $M$ and radius $r$
will have characteristic velocities of the order $\sigV = \sqrt{GM/Cr}$, where
the factor $C$ depends on the geometry and the actual mass distribution, as
well as on whether the system is rotation- or dispersion-dominated
\citep{bin08}.
For dispersion-dominated galaxies, $M$ refers to the total, dynamical mass of
the system, and $C \simeq 6.7$ for various galactic mass distributions
\citep{foe09}. For rotation-dominated galaxies, the appropriate mass is the
mass enclosed within $r$, and $C \simeq 2.25$, again averaged over various
galactic mass distribution models \citep{epi09}.

Measured values of the \HI\ velocity dispersion, or of the stellar velocity
dispersion which arguably reflects the velocity field of the gaseous
components, range from $\sim 10$--$30$ \kms for dwarf galaxies
\citep[e.g.][]{pet93,van98} to several hundred \kms\ for large ellipticals
\citep[e.g.][]{mce95}.  Note that observationally what is measured is the
dispersion along a line of sight, or an average of many lines of sight, and
hence correspond to the dispersion in one dimension. If there is no preferred
direction of motion, the three-dimensional velocity dispersion, which is what
is referred to in \sec{sigVcl_invest}, is a factor $\sqrt{3}$ times higher.

Our fiducial model has a total \HI\ mass of $6.7\times10^8\,\Msun$. For small
galaxies, the ratio of \HI\ to the total, dynamical mass is roughly 0.1
\citep{ski87,van97,van98}. Thus, the velocity dispersion should be
approximately 50 \kms.

At high redshift, most galaxies may not yet have had time to form a disk, and
may thus be expected to be dispersion-dominated. For galaxies which eventually
settle into a disk, the velocity dispersion tends to decrease
\citep[e.g.][]{tho12}. For instance, for intermediate- to late-type spirals,
\citet{ber11} found that the central vertical velocity dispersion is $\sim1/4$
of the maximum rotation speed.

The preceding reasoning dealt with the overall velocity dispersion of the
galaxy's components. For relaxed disks, these could be correlated in phase
space, such that locally to a given photon's emission site the first few clouds
encountered will have a lower relative velocity. With a lower limit of 5 \kms,
however, we believe that a realistic threshold has been met; for instance, in
the MW the dispersions in peculiar motion of local stars with respect to the
local standard of rest is $\sim 20$ \kms\ \citep{sch10}.
%

The velocity
dispersion of the gaseous component might be expected to be smaller than the
stellar velocity dispersion, however, but in The \HI\ Nearby Galaxy Survey
(THINGS), the galaxies, which are almost exclusively disk galaxies, exhibit
velocity dispersions between 10.1 and 24.3 \kms, with a mean of $16.8\pm4.3$
\kms\ \citep{ian12}.


\subsection{Galactic outflow velocity}
\label{sec:Vout_real}

Although not ubiquitous, galactic winds seem to be rather common in
high-redshift galaxies. The outflows are produced by the kinetic and/or thermal
feedback of massive stars and supernovae on the ISM. Radiation pressure and
heating engender expanding bubbles of primarily ionized gas that sweep up
interstellar material, eventually escaping the galaxy (or possibly re-entering
the system after having reached several virial radii) \citep{hec02}. Outflows
are usually detected through low-ionization absorption lines such as
\ion{Mg}{ii} and \ion{Fe}{ii} which appear blueshifted with respect to the
systemic velocity. Typical values are of the order 100 \kms\
\citep[e.g.][]{pet01,rub11}, but range all the way from 0 and up to
$\sim1$--$2000$ \kms \citep[e.g.][]{sha03}.

Various models for galactic outflows have been put forward:
\citet{ors12} equate \Vout with the circular velocity, which is of
the same order as \sigV, discussed in the previous section.
\citet{ber05} and \citet{gar12} apply a weak dependency on SFR, with
$\Vout \propto \textrm{SFR}^{0.145}$, where the SFR is given is
\Msun\ yr$^{-1}$, and the constant of proportionality is $\simeq 300$--1000
\kms.

In the context of LAEs, the effect of an outflow is to diminish the blue peak
of the otherwise symmetric, double-peaked line profile. This is noticable
already $\sim10$ \kms, and by $\sim100$ \kms, the blue peak may be erased
altogether \citep[e.g.][]{dij06a}. Most high-redshift LAE profile observations
have resulted in the well-known, asymmetric red peak only. However, using
sufficiently high resolution, $R \gtrsim 1$--2000, it seems that a significant
fraction (20--50\%) of LAEs actually show at least signatures of a blue peak,
indicating only modest outflow velocities \citep{ven05,tap07,kul12,yam12}.

%
%
%

This hypothesis is in qualitative concord with the recent findings of
\citet{has12}, who interpret the smaller offset of LAE \lya\ lines with respect
to the systemic velocity, compared with those of LBGs, as being due to smaller
outflow velocities. Note, however, that the line offset to a large degree
depends on the column density, which due to the generally much higher gas mass
of LBGs naturally will be larger.

For the reasons given above, and since a large outflow velocity was found to
quicky destroy the boost anyway, we will restrain ourselves to investigating
rather small values of \Vout, from 0 to 100 \kms.


\subsection{Emission sites}
\label{sec:Xinit_real}

Since evidently the extend of the photon-emitting region is not of major
importance (see \sec{Xinit_invest}), for simplicity we will confine our grid of
models to one value only, viz. $H_{\mathrm{em}} = 1$ kpc. This value is roughly
equal to the UV half-light radius of LAEs at redshifts $2\lesssim z \lesssim 6$
\citep[see Fig.~2 of][]{malh12}.

The typical environment from which photons are emitted is more crucial. 
Originating in the \HI/\HII-boundaries surrounding massive stars, the concept
of emitting the \lya\ photons from the ICM is rather dubious. Massive stars are
predominantly born in giant molecular clouds \citep[GMCs; e.g.][]{gar99}
Since such stars are short-lived, with lifetimes of but a
few Myr, they are not expected to travel very far from the dense clouds from
which they were born. This favors a high value of \Pcl, the probability of
being emitted from a cloud rather than from the ICM. On the other hand the
intense UV radiation carves out a Str\"omgren sphere, which reaches a size of
the order of 10(s) pc in a few Myr \citep{hos06,gen12}.
If they are matter-bounded rather than radiation-bounded, the \lya\
photons start to stream freely out into the surrounding, lower-density medium.

Although \citet{isr78} found that most massive stars are located in the
outskirts of GMCs (in the MW), \citet{wal87} argue that they tend to be more
centrally located, but that $\sim30$\% of the \HII\ regions used to probe O and
B stars in reality are just  parcels of ionized gas that was once in molecular
form near the surface of its host cloud. On the other hand, \citet{gen12}
reason that internal turbulence in the GMCs creates filamentary structure so
that any star has a high probability of being born near the surface.

Regardless of the massive stars being born deep in the GMCs, or being born
closer to the edge so that blisters allow a high escape fraction of ionizing
and hence \lya\ photons, the GMCs themselves are usually embedded in the WNM.
The density of
the WNM being several orders of magnitude lower than the GMCs, a Str\"omgren
sphere that breaks out of a GMC may be able to grow faster, faciliting escape
into the ICM.

Massive star do also exist in exposed cluster, such as in the Pleiades.
Although the stars necessarily are born in dense clouds, some stellar groups
manage to ionize and blow away the neutral gas. Observations of this at high
redshifts are difficult, but in the nearest 2 kpc of the Sun, \citet{lad03}
estimate the (lower limit) birthrate of embedded clusters to be 2--4 Myr$^{-1}$
kpc$^{-2}$, which is more than an order of magnitude higher than that of open
clusters \citep[0.25--0.45 Myr$^{-1}$ kpc$^{-2}$][]{elm85,bat91}.

In a study of 45 GMCs in M33, \citet{ima11} found 29 to be spatially and
kinematically coincident with a local peak in atomic gas, 13 to be
kinematically coincident, but located near the edge or on a filament between
two peaks, and the remaining three not to be associated with any high-density
atomic gas. In the region NGC 602/N90 in the SMC, \citet{gou12} found an
``unusually large fraction'' of 60\% of the pre-main sequence stars to be
clustered, while the rest are diffusely distributed in the intercluster area.

Additionally, in the MW 10--20\% of all O stars are found in
ultracompact \HII\ regions, still embedded in their natal molecular cloud
\citep{chu90}.  

From these considerations, we consider $\Pcl = 0.1$ a lower value,
0.2--0.5 to be a more or less realistic value, and 0.9 to be a highest value.


\subsection{ICM dust density}
\label{sec:ZICM_real}

Since the various ways of destroying dust tend to correlate with processes
that also ionize gas (collision, sputtering, sublimation, and evaporation), the
dust-to-metal ratio \xid, and hence the dust-to-gas ratio, in ionized gas may
be expected to be lower than in neutral regions.
However, most extinction curves are obtained from sightlines crossing several
phases of the ISM, although typically only the correlation of the extinction
with the neutral hydrogen is probed. Hence, the variation of \xid\ with \xHI\
is not well-constrained.
Whereas \xid\ in neutral gas seems to be more or less universal over both
different phases, metallicities, and redshifts, the observations
that do exist generally indicate a lower \xid\ in regions of ionized gas,
albeit with large variations.
On the basis of a discussion of various such regions, \citet{lau09b}
argue (Sec.~2.1.1) that \xid\ in these regions ranges between a factor of
\ten{-4} lower than in the neutral phases, to values similar to these, but with
typical values of \ten{-2} times lower.
This is an average over several different types of \ion{H}{ii} regions, in
particular also the very dust-depleted IGM; in the HIM $\xid = \ten{-1}$ may be
more representative. A similar discussion was given recently by
\citet[][Sec.~5]{pal12}, where also it is noted that the mere radiation
pressure may be able to clear large volumes of dust.



\subsection{Emission line width}
\label{sec:sigline_real}

The intrinsic \lya\ line width is given by a convolution of the natural,
thermal, and turbulent broadening. As discussed in \sec{nHIcl_real}, turbulence
may dominate over thermal motions. Since scattering changes the line shape in a
highly non-trivial way, \lya\ line shape observations do not reveal the
intrinsic shape. Instead the width of non-resonant lines such as H$\alpha$,
which are produced in the same locations as \lya, arguably can be used as a
probe. Since the emitting regions also exhibit larger-scale motions, however,
when observing the line integrated over the full galaxy it will usually be much
broader.

When galaxies are well resolved, the internal motion of cloud are more easily
disentangled. In nearby galaxies, \citep{yan94} found H$\alpha$ line widths (in
terms of standard deviations) of 20--30 \kms. At higher redshifts,
using CO lines where large-scale motions have been removed, \citet{swi11}
determine the internal velocity dispersion of the components of a starburst
galaxy at $z = 2.3$ to be 45--85 \kms.


\subsection{Cloud size distribution}
\label{sec:rcl_real}

As mentioned previously, the covering factor rather than the actual shapes and
sizes of the clouds are the crucial factor in determining the boost. However,
for $\Pcl > 0$, the size of the clouds become important, as smaller clouds
implies an easier escape from the first cloud. On the other hand, at a given
covering factor a larger cloud size implies more attenuation of the continuum.
In \sec{rcl_invest} we used $\beta = 2$ when keeping \rcl\ constant in order to
follow \han. Measured slopes tend to be more shallow, $\beta \sim 1.6$
\citep{dic89,wil97}.
Obviously, in reality the WNM does not consist of spherical clouds, so in
order for the conclusions not to depend too much on the chosen cloud sizes, we
will study a broad range of cloud sizes and distributions. Approximately half of
the simulations will be run with fixed cloud size, both $\rcl = 30$ pc and
$\rcl = 100$ pc, while the other half is run with $\rcl = 20$--200 pc and
$\beta = 1$--2.5. This cloud size range is roughly consistent with that of the
LMC \citep{kim03}. Smaller cloud sizes probably requires further cooling, while
larger clouds will get torn apart by large scale motion, at least in
rotation-dominated objects \citep{new80}.



\section{Results}
\label{sec:res}

Having settled upon more or less realistic ranges of the parameters important
for the RT, summarized in \tab{params}, a large number ($4\e{4}$) of
simulations is carried out in the many-dimensional parameter space \P\ spanned
by the parameters.
\begin{deluxetable}{lllll}
\tablecolumns{5}
\tablewidth{0pc}
\tablecaption{Summary of parameter values}
\tablehead{
\colhead{Parameter}    & Full range          & ``Typical''          & ``Unusual''  & Sec.                    \\
}                                                                                                              
\startdata                                                                                                     
\nHIcl                 & 0.03--3             & 0.2--0.5             & 0.1--1       & \ref{sec:nHIcl_real}    \\
$T_{\mathrm{cl}}$      & 5\e{3}--3\e{4}      & 8\,000--20\,000      & Full range   & \ref{sec:nHIcl_real}    \\
$Z$                    & 0.03--2             & 0.05--0.3            & 0.03--1      & \ref{sec:Zcl_real}      \\
\sigV                  & 5-100               & 30--50               & 10--80       & \ref{sec:sigVcl_real}   \\
\fc                    & 0.8--8              & Full range           & Full range   & \ref{sec:fc_real}       \\
\rcl                   & 0.03--0.2           & Full range           & Full range   & \ref{sec:rcl_real}      \\
$\beta$                & 1--2.5              & Full range           & Full range   & \ref{sec:rcl_real}      \\
\nHIICM                & \ten{-12}--\ten{-6} & \ten{-10}--\ten{-7}  & Full range   & \ref{sec:nHIICM_real}   \\
$T_{\mathrm{ICM}}$     & 3\e{5}--5\e{7}      & \ten{5.5}--\ten{6.5} & Full range   & \ref{sec:nHIICM_real}   \\
\Vout                  & 0--100              & 10--50               & Full range   & \ref{sec:Vout_real}     \\
$H_{\mathrm{em}}$      & 1                   & 1                    & 1            & \ref{sec:Xinit_real}    \\
\Pcl                   & 0--1                & 0.2--0.5             & 0.1--0.9     & \ref{sec:Xinit_real}    \\
$\sigma_{\mathrm{em}}$ & 5--100              & 10--85               & Full range   & \ref{sec:sigline_real}  \\
\rgal                  & 5, 10               & 5, 10                & 5, 10        &                         \\
\enddata
\tablecomments{Densities are given in \pcmc, metallicity in terms of the
  Solar value, velocities and line widths in \kms, temperatures in Kelvin,
  distances in kpc.}
\label{tab:params}
\end{deluxetable}
Whereas in all previous simulations \ten{5} photons were used to ensure
convergence, in order to be able to adequately sample \P, we use only $10^3$
photons per model. In most simulations, the resulting boost is accurate to
within roughly 10\%; in simulations where the continuum escape fraction is very
low, usually implying a large boost, this tends to overestimate the boost.
One hundred of such simulations were resimulated to convergence; of these, five
were found to
have underestimated $b$ by 20--30\%, while the rest had overestimated $b$ by
$\sim0$--300\%. However, as will be evident below, all of these models
correspond to physically extremely unlikely models.

\begin{figure*}[!t]
\centering
\includegraphics [width=0.75\textwidth] {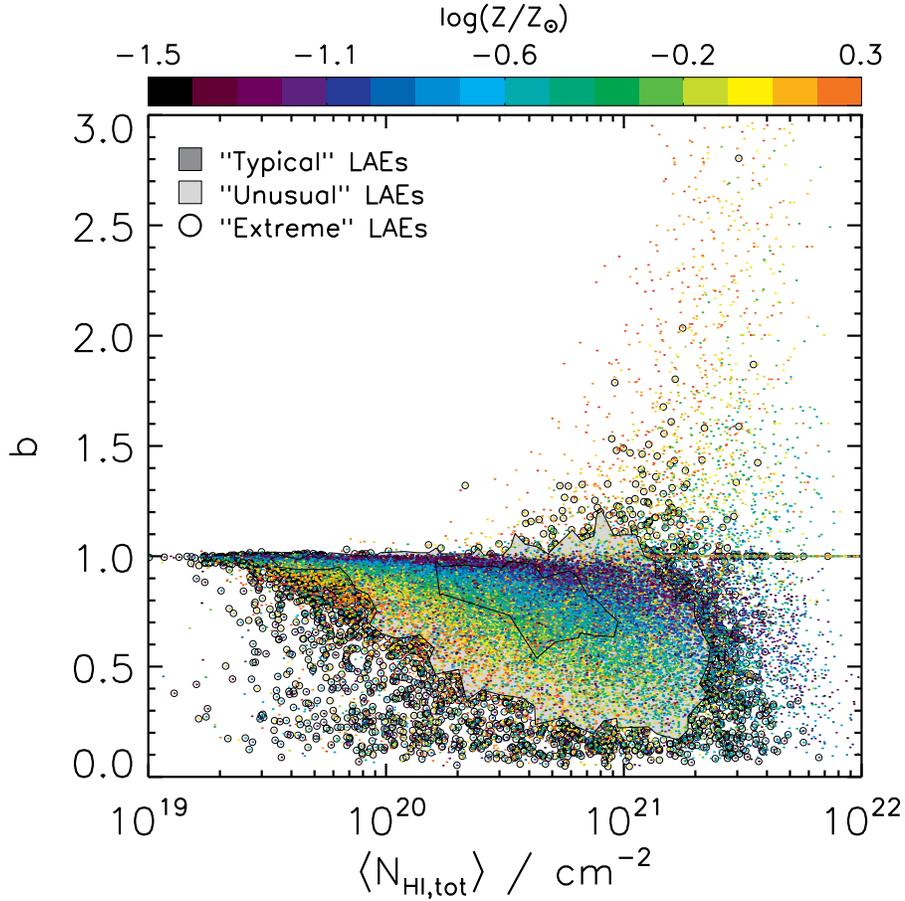}
\caption{``Likelihood map'' of the obtained boosts for models with realistic
  and sub-realistic parameter values. The boost is shown, arbitrarily,
  as a function of average neutral hydrogen column density $\ave{\NHI}$, as
  measured from the center to the surface of a galaxy. Each point represents a
  galaxy, the physical parameters of which are a realization of random values
  in the ranges discussed in Secs.\ \ref{sec:nHIcl_real} through
  \ref{sec:rcl_real} and summarized in \tab{params}.
  The metallicity of a galaxy is depicted by the color of the associated point.
  Models of ``typical'' and ''unusual'' parameters all lie within the
  \emph{dark} and \emph{light gray} regions, respectively. The rest of the
  models have rather unrealistic parameters, but the ones that have either
  $\nHIcl \leq 1.5$ \pcmc, $Z \leq 1.25$, $\sigV \geq 10$ \kms, or $\Pcl \geq
  0.10$ are marked by \emph{black circles}.}
\label{fig:real}
\end{figure*}

The results of the simulations are displayed in \fig{real}, where the boosts
of all models are shown as a function of total, average column density
\ave{\NHI} of neutral hydrogen, colored according to their metallicity.
The gray-shaded contours indicate regions of likelihood, based on the parameter
value ranges listed in column 2 and 3 of \tab{params}. ``Typical LAE'' models
all lie within the dark gray area, while ``unusual'' models lie within the
light gray area.  
%

The remaining models all have rather unrealistic values. We have marked with
black circles the ones for which nevertheless at least one of the following
inequalities is true: $\nHIcl \leq 1.5$ \pcmc, $Z \leq 1.25$, $\sigV \geq 10$,
\emph{or} $\Pcl \geq 0.10$; these models are regarded as 
``extreme, but possibly conceivable'', while the remaining are regarded as
``extreme and probably inconceivable''.


\section{Discussion}
\label{sec:disc}

By far, the majority of the models reveal a boost of $\lesssim 1$. Furthermore,
the ones that do exhibit $b > 1$ tend to have supersolar metallicity, very
high cloud \ion{H}{i} densities ($\sim5$--$10\times$ typical values),
very low velocity fields (both \Vout\ and \sigV\ are $<10$ \kms),
virtually empty
intercloud media, and a high fraction ($> 80$\%) of photons born in the ICM.

Of the models labeled ``extreme, but possibly conceivable'' (indicated in
\fig{real} by the black circles), twelve has resulted in a boost of $b > 1.5$.
Taking a closer look at these model reveals that they all lie close to the
(admittedly somewhat arbitrary) threshold between ``conceivable'' and
``inconceivable''. Moreover, although outflow velocity was not considered as
a defining threshold, all except one have $\Vout \sim 1$ \kms. One model has
$\Vout = 17$ \kms, but has slightly supersolar metallicity ($Z = 1.1 \Zsun$),
low velocity dispersion ($\sigV = 14$ \kms), low emission cloud-correlation
($\Pcl = 0.15$), narrow intrinsic emission line ($\sigma_{\mathrm{em}} = 15$
\kms), high cloud densities ($\nHIcl \simeq 1.5$ \pcmc) while virtually empty
ICM ($\nHIICM \simeq \ten{-10}$ \pcmc).

\begin{figure}[!t]
\centering
\includegraphics [width=0.40\textwidth] {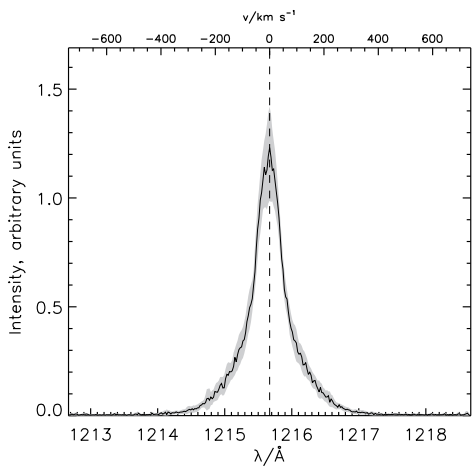}
\caption{Average spectrum (\emph{solid line}) and 1$\sigma$ region
  (\emph{gray-shaded area}) of the marginally realistic models giving a boost
  larger than $b=1.5$. With an average (maximum) FWHM of 110 (150) \kms, these
  lines are much narrower than typical observed high-EW LAEs.}
\label{fig:spectra}
\end{figure}

The twelve models were all resimulated with \ten{5} photons. These
resimulations reduced the value of the boost somewhat in all but one case
(where it increased by a few percent). Furthermore, inspecting the emerging
spectra (\fig{spectra}) shows that \emph{all models exhibit very narrow
emission lines}.
In order for the Neufeld scenario to work, the \lya\ photons by definition are
not allowed to scatter much, lest they
would diffuse too much in frequency, eventually being so far from the line
center that they would be able to penetrate the clouds. The resimulated lines
have, on average, a FWHM of 110 \kms, and the broadest line has a width of only
150 \kms. This is much narrower than observed \lya\ lines, which typically are
many hundreds of \kms\ broad, and rarely below 250--300 \kms\
\citep[e.g.][]{rho03,fyn03,hu04,shi06}. As a further consideration, in spite of
the vanishing outflow velocities, almost all of the spectra lack the prominent
double-peaked feature of \lya\ radiation escaping a static medium.

Additionally, if a given ``too high''-EW LAE is supposed to be caused by a
multiphase medium preferentially absorbing the continuum, rather than some
effect increasing the \lya\ flux, then that LAE should also exhibit a certain
reddening of the continuum. In general, however, the opposite seems to be the
case: In a large sample of COSMOS LBGs and LAEs, \citet{mall12} found a clear
anti-correlation between the ratio of SFRs calculated from the \lya\ and from
SED fitting, respectively, and \ebv.
Similar results were found by \citet{sha03} and \citet{pen09}.
On the scale of individual galaxies, \citet{ate08} found in at least four out
of six nearby galaxies that the regions of high EW correlate with the regions
of low \ebv. Most of the high EW regions is diffuse \lya\ emission, which is
probably \lya\ photons produced in the star-forming regions being scattered
toward the observer.

\subsection{Alternative scenarios}
\label{sec:alt}

Based on the results presented above, it seems that the Neufeld scenario will
allow for an increased \lya\ EW under very special circumstances only, and thus
serves as an very unlikely explanation for the observed high \lya\ EWs.
The reports of high EWs nevertheless exist and warrants an explanation. 
In the following we hence discuss miscellaneous alternatives.

\subsubsection{Measuring errors}
\label{sec:errors}

Firstly, it must be remembered that the mere report of an enhanced EW not
necessarily implies an enhanced EW.
For faint continua,
determining the EW is associated with considerable error, and most of
proclaimed high-EWs have huge error bars.
For instance, \citet{hen10} found six sources with $\ewem > 240$ {\AA}, but
caution that the uncertainties in such high EWs typically exceed 100 {\AA}.


\subsubsection{Stellar population}
\label{sec:pop}

The theoretical upper limit for the \lya\ EW of 240 {\AA} assumes a
\citet{sal55} IMF at Solar metallicity.
For decreasing metallicites, the main sequence of a stellar population is
shifted to the blue, resulting in a slower decline of the ionizing photon
production, and thus a higher \lya\ EW \citep{sch03}. Nevertheless, even for
$Z \sim 0.02\Zsun$, the EW of a starburst is ``only'' 350 {\AA},
i.e.\ corresponds to an interpreted boost of $\sim1.5$.

For even lower values of $Z$ very high values of \ewint\ may be reached.
An extreme case of a low-metallicity population would be one consisting of
PopIII stars \citep[e.g.][]{tum03,zac11}. While still not observationally
confirmed, stars born from (close to) metal-free gas are expected to be able to
reach very high masses. Being particularly luminous in the Lyman continuum,
such a population of stars will be able to emit a remarkable fraction of their
photons in \lya. Whereas a ``normal'' population emits 6--7\% of its bolometric
luminosity in \lya, 10--40\% can be reached for sufficiently low metallicities,
with a resulting \ewint\ of several thousands of {\AA}ngstr\"om
\citep{bro01,sch02,rai10}. However, at metallicities this low, no dust
should yet have formed, and thus no reddening should be observed. Moreover,
with such a hard spectrum, the EW of the \ion{He}{ii} H$\alpha$ line at 1640
{\AA} is expected to be noticable. In some of the high \lya-EW cases, this was
specifically looked for, but not detected \citep{daw04}.

Shifting one or both of the IMF mass limits toward higher masses, or applying a
shallower slope, implies a relatively larger fraction of high-mass stars, with
a harder ionizing UV spectrum and thus a larger \ewint. For instance,
increasing the upper mass limit from 100 to 500 \Msun\ (corresponding to a rise
in average mass from 3.1 to 3.4 \Msun\ only) for a $Z = 5\times10^{-4}\Zsun$
population, strengthens \ewint\ from $\sim400$ {\AA} to $\sim700$ {\AA}, while
further increasing the lower mass limit from 1 to 500 \Msun\ (corresponding to
a rise in \ave{M} to 112 \Msun) strengthens \ewint\ to $\sim900$ {\AA}
\citep[see Fig.~7 of][]{sch03}.

In all cases, the above values assume an instantaneous burst of stars; with the
early O and B star dying fast, the EW of these extreme populations quickly
decline, reaching more modest values after only a few Myr, and would thus have
to be observed in a special period of their lives. For continuous star
formation, high EWs can be found at any point in time, although the values will
always be lower than the initial EWs of the instantaneous bursts. For $Z =
5\times10^{-4}\Zsun$, \ewint\ increases from $\sim180$ {\AA} to $\sim240$
($\sim500$) {\AA} for a mass range 1--500 (50--500) \Msun. This can be
compared to $\ewint\sim100$ {\AA} or less for $Z\geq0.02$ \Zsun.


\subsubsection{Delayed escape of \lya}
\label{sec:delay}
In principle, for a short-lived starburst the \lya\ photons --- having their
path length out of the galaxy increased --- could be observed after the
continuum has already faded, resulting in a higher EW. This was investigated
for a homogeneous sphere of dust-free gas by \citet{roy10} and \citet{xu11},
who found that, depending on the column densities, the \lya\ radiation
may be appreciably delayed.
When clumpiness is introduced, we find that the escape time decreases
significantly. For instance, if all the gas in the fiducial model is
distributed homogeneously, the \lya\ photons escape with a typical delay of
$t_{\mathrm{esc}}\sim100$ kyr if there were no dust. In contrast, with the
clumpy structure, $t_{\mathrm{esc}}\sim20$ kyr (not considering the photons
that escape directly without interacting with any clouds). When dust is added,
$t_{\mathrm{esc}}$ is reduced further, since the photons that have the longest
escape time are the ones that are most prone to absorption.

For larger $r_{\mathrm{gal}}$, $t_{\mathrm{esc}}$ increases. However, even
among the models with $r_{\mathrm{gal}} = 10$ kpc, few have $t_{\mathrm{esc}}
> 100$ kyr. Compared to the typical lifetime of the O and B stars of several
Myr, a stellar continuum should still be visible as the \lya\ radiation
reaches its maximum.

Solving simultanenously the RT for \lya\ and ionizing UV radiaton in a
spherically symmetric model galaxy, \citep{yaj12c} found that most of the \lya\
photons could remain trapped until the galaxy has been fully ionized, at which
point the ``old'' and the ``new'' \lya\ photons are released together,
resulting in \lya\ EWs of up to $\sim 1000$ {\AA}.
It is unclear, however, how applicable these results are to more realistically
simulated galaxies, where low-density regions are ionized more rapidly,
allowing the \lya\ photons to leak in those directions.

In a cosmologically simulated galaxy of approximate radius 25 kpc, \citet{lau07}
found a typical path length of the \lya\ photons of $\sim 40$ kpc, i.e.\ a
delay with respect to the continuum of $\sim 50$ kyr. This simulation, however,
did not include ionizing UV RT.


\subsubsection{AGN activity}
\label{sec:AGN}

If a galaxy contains an AGN, the \lya\ EW may be much higher than the 240
{\AA}. The line widths of these object are so broad, however, that such objects
can usually be excluded, as in the case of e.g.\ \citet{rho03} and
\citet{daw04}. Even type II AGNs, i.e.\ objects where the broad-line region is
not visible, typically have larger line widths than the observed high-EW
objects. Furthermore, if the majority of the high-EW systems were due to AGN
activity, it would be inconsistent with the X-ray emission
\citep[e.g.][]{wan04,gaw06}.


\subsubsection{Viewing angle}
\label{sec:angle}

\begin{figure}[!t]
\centering
\includegraphics [width=0.40\textwidth] {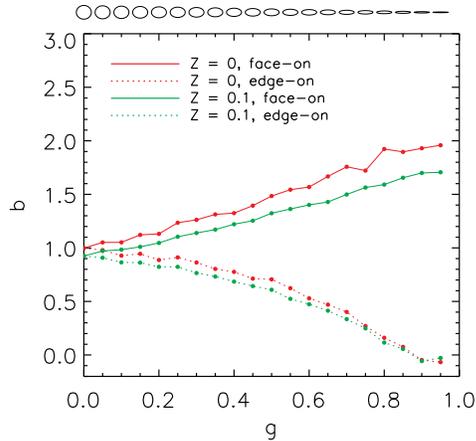}
\caption{EW ``boost'' $b$ when observing the system face-on (\emph{solid})
  and edge-on (\emph{dotted}) as a function of flattening \mbox{$g \equiv
  1-c/a$}, where $a$ and $c$ are the semi-major and -minor axes of the system
  (shown schematically as small ellipses in the top of the plot). Results are
  shown for both dust-free (\emph{red}) and dusty (\emph{green}, $Z=0.1\Zsun$)
  gas. In both cases the gas is homogeneous, and the column density is
  $\NHI = 2\times10^{20}$ \pcms, as measured along the $c$-axis. For reference,
  the Milky Way has a flatness of $g\sim0.95$-0.99 \citep{bin98}.}
\label{fig:flat}
\end{figure}

All the models studied in this work are spherically symmetric. For a flattened
system, photons escape more easily face-on than edge-on.
But since \lya\ photons scatter and thus wander around in the galaxy, their
probability of escaping from the face before reaching the edge is increased,
and thus a higher EW will be observed face-on. Without the information of the
system's morphology, this can be interpreted as an EW boost. That this is true
even for a \emph{homogeneous, dust-free} ISM is seen in \fig{flat}, where the
impact on the EW of gradually flattening a sphere of total \ion{H}{i} column
density $\NHI = 2\times10^{20}$ \pcms\ (corresponding to a DLA) is shown.

This effect is of course observationally impracticable, except
statistically, but has also been investigated in more realistic galaxy
simulations. In \citet{lau09b}, \lya\ RT was carried out in a sample of nine
randomly oriented galaxies extracted from a fully cosmological simulation and
resimulated at high resolution. The ratio of the fluxes escaping in the most
luminous to the least luminous direction, roughly corresponding to what would
be interpreted as an EW boost, was in the range 1.5--4 if integrating over the
whole galaxy, and $\sim10$ if looking at the region of maximum surface
brightness only.
From \fig{flat}, this corresponds roughly to ``boosts'' of 1.1--1.5 and
$\sim2$, respectively.
Similar values was found by \citet{zhe10} (a ratio of seven for the central
region of a single galaxy) and \citet{bar11} (a ratio of 1.7--3 for various
implementations of galactic winds). More recently, \citet{yaj12b} found no
preferred direction of escape for a single, cosmologically simulated galaxy at
$z = 3.1$, but a strong dependence on viewing angle at $z = 0$ when the galaxy
had settled into a disk, similar to what was found by \citet{ver12} for an
isolated disk galaxy.


\subsubsection{Cooling radiation}
\label{sec:cool}

At high redshifts, a large fraction of galaxies may be expected to be in the
process of forming, i.e.\ to accrete gas from the surrounding medium, resulting
in collisionally excited \ion{H}{i} leading to \lya\ emission. Since
the implied densities are rather high, the timescale for cooling is generally
smaller than the dynamical timescale, and thus the infalling gas should be
relatively cold \citep[$T\sim10^4$ K, e.g.][]{fal85,hai01}. Consequently,
roughly 50\% of the energy is emitted in the \lya\ line alone \citep{far01}.

For relatively massive galaxies, \citet{goe10} found at $z = 2.5$ that cold gas
accreting can give rise to significant \lya\ emission of up to a few times
$10^{-17}$\ergscm2~arcsec$^{-2}$ on scales of 10-100 kpc, which is comparable
to typical line strengths.
Such gravitational cooling not only gives rise to \lya\ radiation but also to
a continuum, although relatively it is much fainter than the stellar continuum.
\citet{dij09a} calculated analytically the EW of the cooling radiation from a
collapsing halo and found that, depending on where the continuum is measured,
the intrinsic EW of the cooling radiation is $\gtrsim950$ {\AA}.
Numerically, a few studies have addressed this issue and find that in general
the fraction of the \lya\ radiation that stems from gravitational cooling
increases with $z$,
from $\sim10$\% at $z=3.6$ \citep{lau07},
to $\sim16$-18\% at $z=5.7$--6.5 \citep{day10},
to $\sim50$-60\% at $z=7$--8 \citep{jen12}. Similarly, \citet{yaj12a} found that
cooling radiation becomes comparable to recombination \lya\ for $z > 6$.
Obviously, large deviations from these values exist, both numerically and in
reality, as in the case of \lya\ blobs where in some cases no stellar
continuum is seen at all; a liable explanation for these objects is cold
accretion \citep{nil06,dij09b}

However, while the photons of stellar origin, both \lya\ and continuum, tend to
be born in regions of comparatively high densities of both dust and neutral
gas, thus being more prone to dust absorption, the cooling photons may escape
more or less freely. Thus, cooling radiation could dominate even at quite low
redshifts, although the actual fraction is difficult to assess observationally
as long as the escape fraction of stellarly produced photons is still
associated with so large uncertainties.


\subsubsection{IGM extinction}
\label{sec:IGM}

At high redshifts, neutral hydrogen in the IGM scatter a significant fraction
of the photons blueward of the \lya\ line out of the line of sight. This
affects both the continuum and the line itself, and is sometimes accounted for
using a model of the IGM absorption. For instance, \citet{mal02} use the
prescription of \citet{mad95} and found that the corrections for the broadband
and the narrowband approximately cancel, leaving the EW unaffected. This
assumes that the \lya\ line is symmetric about the line center; if the red
peak is enhanced relative to the blue peak, as would be the case for an
expanding medium, the IGM transmits more \lya, resulting in the erroneous
interpretation that the EW is larger than in reality. Moreover, as the galaxies
detected in \lya\ at high redshifts may be biased toward being the ones that
have made themselves visible by ionizing a bubble around them in an otherwise
partly neutral IGM, the line may be less affected by the IGM than the
continuum, which again would be interpreted as an EW boost.

Even if the correlation of the state of the IGM with the sources is taken into
account, large fluctuations in the neutral fraction of hydrogen exists, which
may also be a possible source of error \citep[e.g.][]{lau11,jee12}. Using a
simple model for LAEs at high redshifts
but incorporating clumpiness of the IGM, \citet{hay06} calculate distribution
functions for observed EWs, for various narrowband/broadband techniques. They
found that, depending on the redshift, the median \ewobs\ may easily lie
appreciatively above \ewem, and that for instance at $z = 5.7$ with narrowband
and broadband both centered on the \lya\ line, 20\% of the EWs will be boosted
by a factor of 2 or higher.  


\subsubsection{Star formation stochasticity}
\label{sec:SFstoch}

When calculating the resulting EW from a distribution of stellar masses, an
infinite number of stars are usually assumed.
Using the numerical code {\sc SLUG} \citep{fum11,das12}, \citet{for12} found 
that for low-SFRs galaxies, stochasticity alone results in a distribution
of EWs from $W_0/4$ to $3W_0$, where $W_0$ is the mean EW. However, since this
effect is minimal for SFRs $\gtrsim 1$ \Msun\ yr$^{-1}$, which are hardly
visible at high redshifts, it cannot serve as an explanation for most of the
observed high-EW galaxies.


\subsubsection{Inhomogeneous escape}
\label{sec:inhomoesc}

For resolved galaxies, regions with no star formation may scatter \lya\ from
distant regions toward the observer, such that these particular regions
exhibit extreme EWs \citep{hay07}. Integrated over the source, however,
\lya-overluminous regions cancel out with underluminous regions (modulo other
effects affecting the RT).




\section{Summary and conclusion}
\label{sec:sum}

We have carried out a comprehensive set of \lya\ and FUV continuum radiative
transfer calculations, systematically varying the significant physical
parameters, first separately and then in unison. The aim was to investigate
whether a multiphase interstellar medium is capable of preferentially absorbing
the continuum so as to enhance the observed \lya\ equivalent width. The
motivation for this study are the numerous observations of \lya\ emitting
galaxies exhibiting EWs larger than what is theoretically expected, together
with the previously proposed explanation of such a ``boost'' being a
consequence of a clumpy ISM \citep{neu91,han06}. In the suggested model, a
galaxy consists of a number of clouds, physically corresponding to the ISM
phase termed the ``warm neutral medium'', dispersed in a low-density medium
corresponding to the phase called the ``hot ionized medium''.

We find that, while indeed it is possible to construct a model galaxy with the
said ability, the physical properties needed are extremely unlikely to be found
in a real galaxy. In order for a galaxy to enhance its intrinsic \lya\ EW, a
large number of criteria must be met:
\begin{enumerate}
\item The metallicity must be high ($\gtrsim$ Solar),
\item the density of the WNM must be very high ($\gtrsim 5$ times typical
      values),
\item the density of the HIM \emph{and} its ionization fraction must be very
      low, so that the total density of neutral hydrogen in this phase does not
      exceed $\sim\ten{-7}$ \pcmc,
\item the bulk ($>80$--90\%) of the \lya\ and FUV photons must originate from
      regions where the stars have blown away the neutral gas from which they
      were born,
\item the galaxy must have virtually no outflows ($\Vout \lesssim
      10$ \kms), and
\item the velocity dispersion of the gas clouds must be very low ($\lesssim 10$
      \kms).
\end{enumerate}
In particular the latter two points are important; the whole point of the
proposed multiphase model is that the paths of the \lya\ photons are confined
to a dust- and gasless medium, never penetrating the dusty clouds but instead
scattering off of their surfaces. This mechanism works because of the resonance
nature of \lya\ scattering. However, as soon as the clouds have a small
velocity, especially random motions, the \lya\ photons are shifted out of
resonance in the reference frame of the clouds, allowing them to penetrate
much farther into the clouds and thus exposing them to a higher column density
of dust.
%
Note that in principle it \emph{is} possible to have an enhanced EW for larger
velocity fields, but this will require even more extreme conditions regarding
the other parameters.
With regards to the required densities of neutral hydrogen, these are more
characteristic of the so-called cold neutral medium or molecular clouds.
Since the volume occupied by these phases are much smaller than that of
the WNM \citep[e.g.][]{fer01}, the covering factor of these phases is small.
Moreover, they will typically be embedded \emph{inside} the WNM, affecting
only those photons that were already trapped in the clouds.
In addition to the points mentioned above, it must also be remembered that the
\emph{intrinsic} \lya\ EW of a galaxy exhibiting a high observed EW must be
high, i.e.\ the stellar population must be very young ($\sim$a few Myr);
otherwise an even higher boost would be needed to bring the EW above the
theoretical limit.

In conclusion, we consider the Neufeld model to be an extremely unlikely
reason for the observed high EWs. We have discussed a number of other
possible explanation (\sec{alt}), of which the most probable, in our opinion,
is excess of \lya\ radiation from cold accretion (may contribute $\sim 0.1$--1
times the stellar \lya, depending on redshift, but may escape much more
easily), and/or anisotropic escape of \lya\ (the \lya\ flux may vary in
different directions by a factor $\sim1$--10, depending on irregularity of the
galaxy and aperture of the detector.

\acknowledgments
We are grateful to Jens Knude, Sangeeta Malhotra, and Andrei Mesinger for
valuable discussions about the properties of the interstellar medium.
PL acknowledges fundings from the Villum Foundation.
The simulations were performed on the facilities provided by the Danish Center
for Scientific Computing.


\appendix

\section{Extending and testing \moca}
\label{sec:moca}

\subsection{Implementing continuum radiation}
\label{sec:cont}

The original version of \moca\ included only the RT of \lya, i.e.\ the initial
frequency probability distribution of photons followed a Voigt profile with the
Gaussian core given by the temperature/turbulence of the gas. To calculate EWs,
the wavelength range $\Delta\lambda$ transferred though the medium must be
extended so as to include the continuum. For a given EW this is readily
accomplished by first determining whether an emitted photon is a line rather
than a continuum photon, with a probability $P(\textrm{line}) \propto \ewint /
\Delta\lambda$, and then determining the exact wavelength of the photon
depending on the intrinsic line profile or the slope of the continuum (in this
study taken to be flat in $\nu$). For a broad wavelength region around \lya,
numerically there is no difference in the RT of the two kinds of photons.


\subsection{Building the grid}
\label{sec:grid}

In order to construct spherical clouds, we make use of adaptive mesh refinement
(AMR), where cells may be refined into eight subcells. As the clouds themselves
are uniform, high resolution is only needed at their surfaces, not their
interior. To build the AMR grid, we use the following approach:

Given a number $N_{\mathrm{cl}}$ of clouds and a distribution of sizes
specified by a minimum and maximum radius $r_{\mathrm{min}}$ and
$r_{\mathrm{max}}$, as well as a size distribution power law index $\beta$,
cloud centers are placed randomly in a sphere of radius \rgal,
such that no clouds overlap. The clouds' surfaces are then marked by a number
of auxiliary particles, typically $\sim10^3$. Subsequently, the grid is
constructed starting from a mother grid of base resolution $2^3$, recursively
refining any cell that contains more than one particle. Finally, each cell is
labeled either a ``cloud'' cell or an ``ICM'' cell, depending on the distance
to the nearest cloud center, and the physical parameters
then assigned accordingly. Note that all cells belonging to a given cloud share
the same bulk velocity, such that there are no velocity gradients inside a
cloud.

The auxiliary particles can be positioned in different ways so as to constitute
the cloud surface; the most straightforward way is to place them at a random
position at a distance \rcl\ from the center of the cloud. However, it turns
out that in general some particles end up so close to each other, that in order
to refine the cells sufficiently, hundreds of refinement levels may needed.
Distributing the particles evenly on the surface ensures that roughly the same
level of refinement is needed everywhere. Various algorithms exist for this
geometrical exercise \citep[see, e.g.,][]{saf97}.


\subsection{Altering the acceleration scheme}
\label{sec:acc}

To speed up calculations, \moca\ makes use of various acceleration schemes. The
most efficient is the \emph{core-skipping} scheme: Since scatterings of photons
near the line center are usually associated with negligible spatial movement,
while it typically takes of the order $10^5$ scatterings to ``escape'' the
core, these scatterings can be skipped by drawing the random velocity of the
scattering atom from a centrally-truncated Gaussian, favoring fast moving atoms
that can give the photon a large Doppler shift. In \moca, the critical
wavelength defining when the acceleration scheme can be invoked is dependent on
the product $a\tau_0$, such that for higher \HI\ column densities and lower
temperatures, a larger critical value can be used, allowing for a larger
speed-up. The value of $a\tau_0$ is specific to a given cell, and is calculated
from the center of the cell to the face.

Since the code has previously been used for galaxies extracted from
cosmological simulations, cells are usually surrounded by other cells of
similar physical conditions, i.e.\ there are no large gradients. In the
present, idealized simulations, however, where densities may go from zero to
some extreme value over the course of a single cell step, this poses a problem
for photons close to the line center scattering on the surface of a cloud.

In reality, the photon typically escapes after a handful of scatterings,
changing its frequency only of the order 1.5 Doppler widths. With the
acceleration scheme, the first scattering pushes the photon artificially far
out in the wing, rendering the optical depth of hydrogen significantly reduced,
such that the photon is able to penetrate the cloud (or the next cloud) too
much, with a corresponding higher probability of being absorbed by dust.

We solve this issue by calculating $a\tau_0$ not from the cell center, but from
the current position of the photon to the face of the cell. Unfortunately, the
consequence is a quite less efficient acceleration.


\subsection{Testing the code}
\label{sec:test}

\begin{figure}[!t]
\centering
\includegraphics [width=0.33\textwidth] {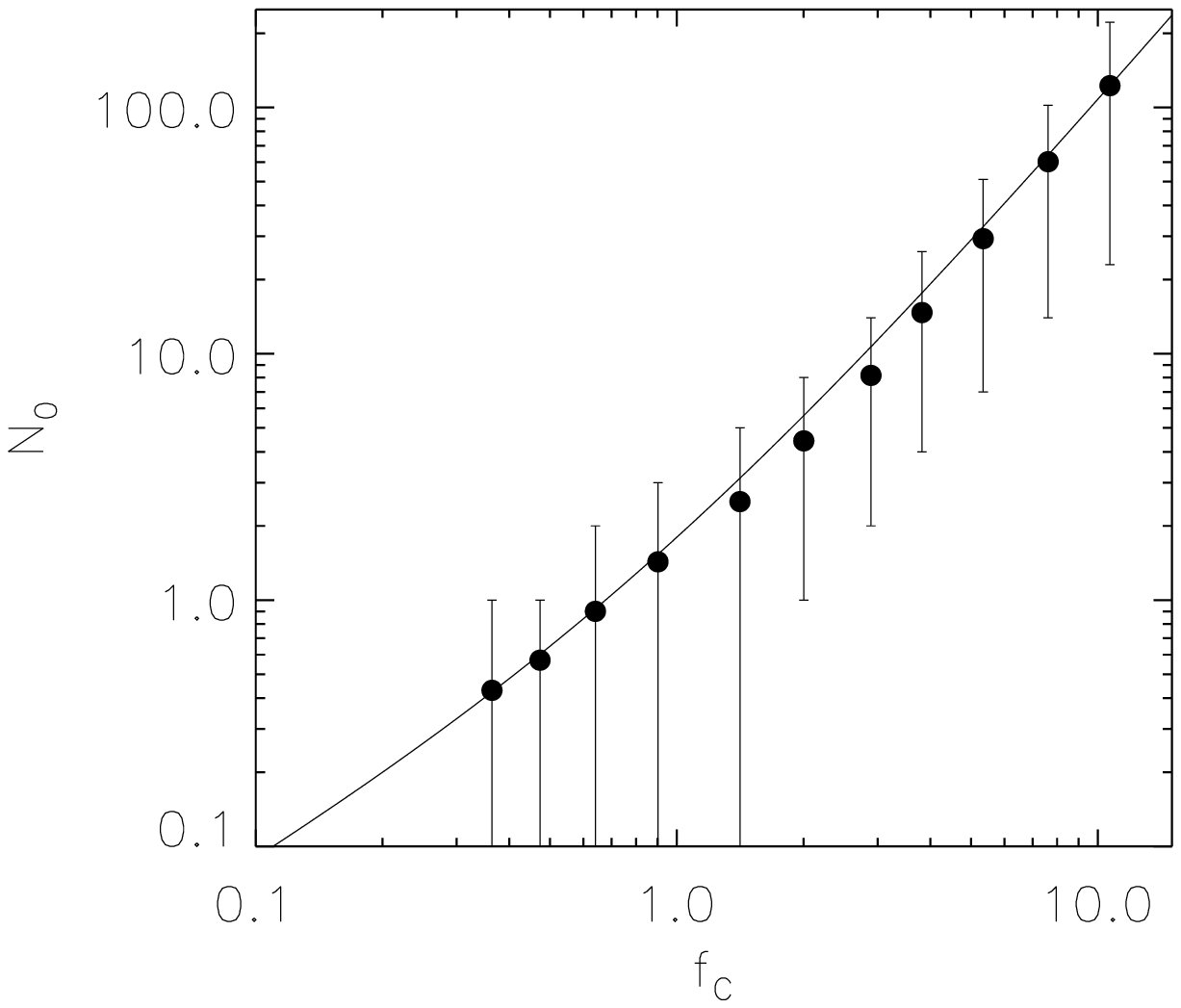}
\includegraphics [width=0.33\textwidth] {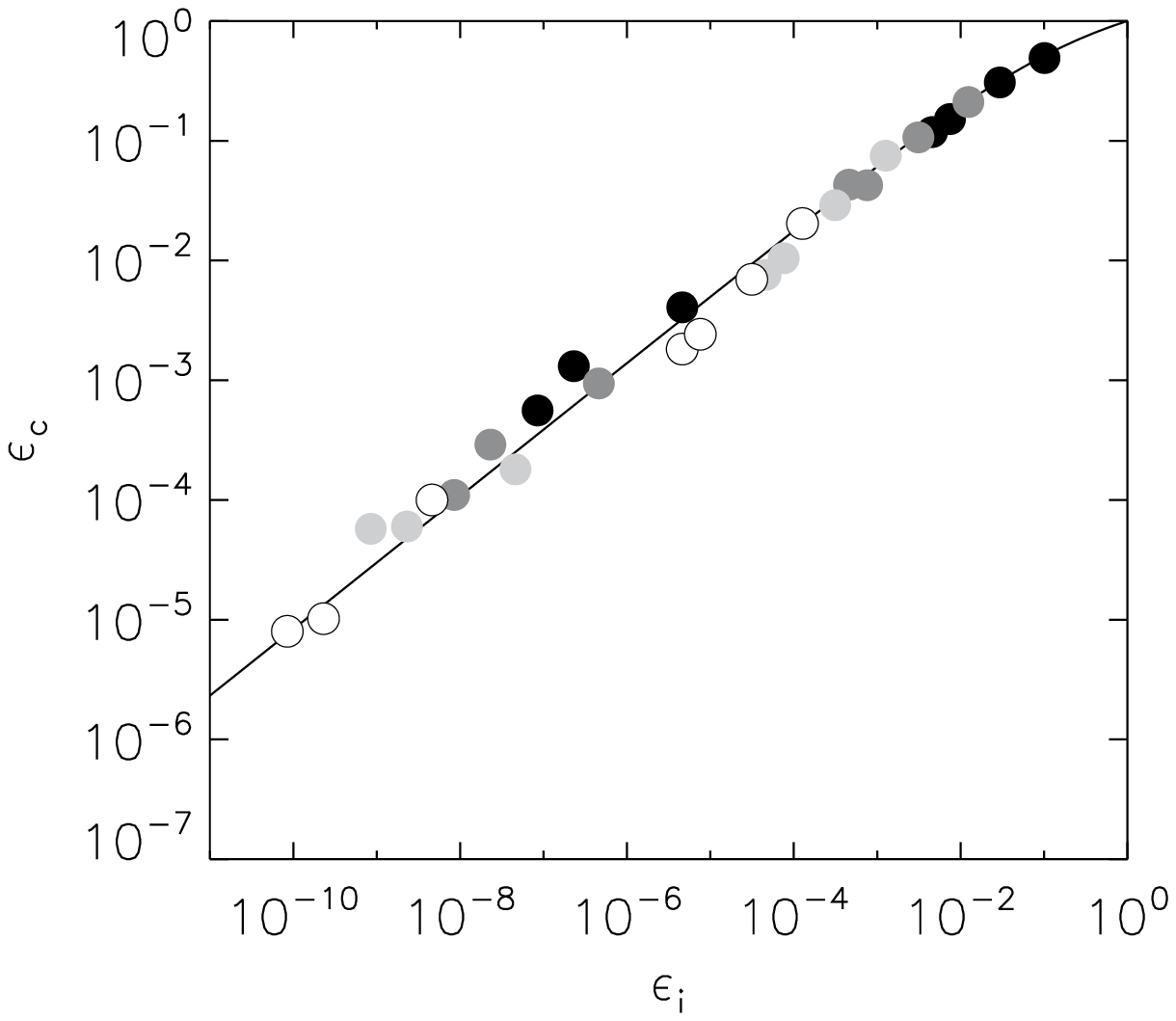}
\caption{{\small \emph{Left panel}: Average number $N_0$ of clouds with which a
  \lya\ photon interacts before escaping from the center of a spherical
  multiphase medium, as a function of the covering factor \fc\ of clouds.
  Error bars indicate the region within which 68\% of the photons fall.
  \emph{Right panel}: Probability of being absorbed rather than reflected from
  the surface of a dusty cloud after numerous scatterings, as a function of
  probability of absorption in a single $[$gas or dust$]$ interaction event for
  a photon at an incident frequency $x_i$ Doppler widths from the line center.
  Each shade correspond to different values of
  dust cross section: From lightest to darkest 0.01, 0.1, 1, and 10 $\e{-21}$
  cm$^{2}$. For each value, seven different incident photon frequencies $x_i$
  were run: From left to right, $x_i = 20, 10, 5, 4, 2, 1$, and 0.
  In both panels, the solid line shows the fits found by \han.
 }}
\label{fig:test}
\end{figure}

In the original Neufeld scenario, that is with no velocity fields etc., the
physical parameters most important to the RT are $N_0$ and
$\epsilon_{\mathrm{c}}$, i.e.\ the average number of clouds with which a
\lya\ photon emitted from the center of a spherical conglomeration of clouds
interacts, and the probability per cloud interaction that a photon (\lya\ or
continuum) is absorbed.
No analytical solution exist for these quantities, but \han\ found that they
could be well fitted by
$N_0 = \fc^2 + \frac{4}{5}\fc$ (in the absence of absorption)
and
$\epsilon_{\mathrm{c}} \simeq 3\epsilon_i^{5/9} / (1+2\epsilon_i^{1/2})$,
where $\epsilon_i$ is the probability of absorption per $[$gas or dust$]$
interaction for an incident photon lying at $x_i$ Doppler widths from the line
center. The results are shown in \fig{test}.
%



\end{document}